\documentclass[journal,onecolumn,11pt]{IEEEtran}
\usepackage{amsthm}
\theoremstyle{plain}
\newtheorem{theorem}{Theorem}
\newtheorem{lemma}{Lemma}
\newtheorem{proposition}{Proposition}

\theoremstyle{definition}

\usepackage{subfig}
\usepackage{setspace}
\def\bm{{\mathbf m}}
\def\bp{{\mathbf p}}
\def\bx{{\mathbf x}}

\def\bP{{\mathbf P}}

\def\cU{{\mathcal U}}
\def\cS{{\mathcal S}}
\def\cN{{\mathcal N}}

\usepackage[cmex10]{amsmath}
\usepackage{amssymb}
\usepackage{array}

\usepackage{algorithm}
\usepackage{algorithmic}
\usepackage{setspace}
\usepackage{epstopdf}
\usepackage[export]{adjustbox}
\usepackage{subfig}
\usepackage{graphicx}
\usepackage{stfloats}
\usepackage{filecontents,lipsum}
\usepackage[noadjust]{cite}

\ifCLASSINFOpdf
\else
\fi

\begin{document}
\title{Optimal Forwarding in Opportunistic Delay Tolerant Networks with Meeting Rate Estimations}

\author{Shohreh Shaghaghian 
        and~Mark Coates
\thanks{The authors are with the Computer Networks Research Laboratory, Department of Electrical and Computer Engineering, McGill University, Montreal, QC H3A 0E9, Canada (e-mail:
shohreh.shaghaghian@mail.mcgill.ca;
mark.coates@mcgill.ca). This research was supported by the Natural
Sciences and Engineering Research Council of Canada (NSERC) via grant
application 261528.}
\thanks{Manuscript received April 3, 2015}}

\markboth{}
{Shell \MakeLowercase{\textit{et al.}}: Bare Demo of IEEEtran.cls for Journals}

\maketitle

\begin{abstract}
  Data transfer in opportunistic Delay Tolerant Networks (DTNs) must
  rely on unscheduled sporadic meetings between nodes. The main
  challenge in these networks is to develop a mechanism based on which
  nodes can learn to make nearly optimal forwarding decision rules despite having
  no a-priori knowledge of the network topology. The forwarding
  mechanism should ideally result in a high delivery probability, low
  average latency and efficient usage of the network resources. 
  In this paper, we propose both centralized and
  decentralized single-copy message forwarding algorithms that, under relatively
  strong assumptions about the network’s behaviour, minimize the
  expected latencies from any node in the network to a particular
  destination. After proving the optimality of our proposed
  algorithms, we develop a decentralized algorithm that involves a
  recursive maximum likelihood procedure to estimate the meeting
  rates.  We confirm the improvement that our proposed algorithms make
  in the system performance through numerical simulations on datasets
  from synthetic and real-world opportunistic networks.
\end{abstract}


\IEEEpeerreviewmaketitle

\section{Introduction}\label{sec:intro}
\IEEEPARstart{D}{\textit{elay}} or \textit{Disruption Tolerant Networks}
(DTNs) are a class of wireless mobile node networks in which the
communication path between any pair of nodes is frequently
unavailable. Nodes are thus only intermittently connected. DTNs were
first studied in the 1990s when the research community considered how
the Internet could be adapted for space
communications~\cite{voyiatzis2012survey}. Later, it was recognized
that DTNs were a suitable model for several terrestrial networks.

DTNs can be categorized according to whether the
node connections are scheduled (thus predictable) or random (hence
unpredictable). Space communication networks fall into the first
group. Networks belonging to the second category, which are the focus
of this paper, are also referred to as {\em opportunistic networks},
because nodes seize the opportunity to transfer data when a
communication channel becomes available. Opportunistic networks have been studied intensively in
recent years (e.g.,
\cite{khabbaz2012disruption,cao2013routing,wei2014survey}) because
they can fulfil a number of useful purposes, such as non-intrusive
wildlife tracking (e.g., ZebraNet~\cite{juang2002energy} and
SWIM~\cite{small2003shared}), emergency response in disaster scenarios
(e.g., ChaosFIRE~\cite{pataki2014sensor}), provision of data
communication to remote and rural areas (e.g.,
DakNet~\cite{pentland2004daknet}), and traffic offloading in cellular
networks (e.g., \cite{han2012mobile}).

In most opportunistic networks, the nodes are highly mobile and have a
short radio range, and the density of nodes is low. In many cases,
nodes have limited power and memory resources. These attributes
combine with the intermittent connections to make routing
traffic challenging. Routing is usually based on a store-carry-forward mechanism that
exploits node mobility. In this mechanism, the source transmits its
message to a node it meets. This intermediate node stores and then
carries the received message until it meets another node to which it
can forward the message. This process is repeated until the message
reaches its destination. The key ingredients in designing an
opportunistic network routing protocol are the forwarding decisions:
should a node forward a message to a neighbour it meets?  should it
retain a copy for itself?

Although much research effort has been devoted to the development of
opportunistic network routing
algorithms~\cite{vahdat2000epidemic,grossglauser2001mobility,spyropoulos2005spray,
  davis2001wearable,lindgren2003probabilistic,burgess2006maxprop,
  jones2007practical,daly2007social,hui2011bubble,sharma2013contact,
  xiao2013community,li2011impact,sermpezis2014understanding,zhang2013gossip,
  conan2008fixed,xiao2013tour,boldrini2010modelling,boldrini2012performance},
the algorithms are either centralized, have no performance guarantees,
or ignore the need to estimate network parameters. Our work focuses on the mobile ad-hoc network (MANET)
setting, where node speed is much reduced compared to the vehicular
ad-hoc (VANET) case, and we can assume that there are fewer
restrictions on the amount of data that nodes can transfer when they meet. In this paper, we
derive a decentralized routing algorithm that has performance
guarantees (under simplifying assumptions about the network
behaviour). When the meeting times between nodes are
independent and exponentially distributed, the routing algorithm
minimizes the expected latency in sending a packet from any source
node to a specific destination. We examine the behaviour of the
routing algorithm when the meeting rates are learned online using a
recursive maximum likelihood procedure. We show that, for a stationary
network, the decision rules and achieved expected latencies converge
to those obtained when there is exact knowledge of the meeting
rates. We present the results of simulations that compare the
performance of the proposed algorithm to previous approaches, and
examine how the algorithm is affected by practical network limitations
(finite buffers, restrictions on data exchange, message expiry times).

{\em Organization}: The paper is organized as follows. In the following
subsection, we discuss related work. In
Section~\ref{sec:system_model}, we describe our system model and
formulate the routing problem. In Section~\ref{sec:algo}, we
present the forwarding algorithms and discuss their
optimality under the network modeling assumptions. We present
numerical simulation results in Section~\ref{sec:res} and make
concluding remarks in Section~\ref{sec:conc}.

\subsection{Related Work}

The first proposed approaches for routing in opportunistic networks
were based on extending the concept of flooding to intermittently
connected mobile networks. In these {\em replication-based} methods, a
node forwards the messages stored in its buffer to all of (or to a
fraction of) the nodes it encounters. There is no attempt to evaluate
the capability of a given node to expedite the delivery. These routing
algorithms have few parameters: they determine only how much
replication can occur and which nodes can make copies of packets. One
of the earliest algorithms was {\em epidemic
  routing}~\cite{vahdat2000epidemic}, in which a node forwards a
message to any node it meets, provided that node has not previously
received a copy of the message. Thus messages are quickly distributed
through the connected portions of the network. Other
replication-based approaches (\cite{grossglauser2001mobility,
  spyropoulos2005spray,ramanathan2007prioritized,khouzani2012optimal})
manage to reduce the transmission overhead of epidemic routing and
improve its delivery performance through modification of the
replication process and prioritization of messages. Models have been
developed that allow an analytical characterization of the performance
of the epidemic routing techniques~\cite{klein2010reaction,wang2012analytical}.
Replication-based approaches result in a high probability of
message delivery since more nodes have a copy of each message, but
they can produce network congestion.

A step towards achieving more efficient routing approaches is to
consider the history of node contacts in the network instead of
blindly forwarding packets. {\em History-based} (also called {\em
  utility-based}) routing algorithms assume that nodes' movement
patterns are not completely random and that future contacts depend on
the frequency and duration of past encounters. Based on these past
observations, both the source and the intermediate nodes decide
whether to forward a message to nodes they encounter or to store it
and wait for a better opportunity. An early example is
\cite{davis2001wearable}, which extends epidemic routing to situations
with limited resources, incorporating a dropping strategy for the case when the buffer
of a node is full. The dropping decisions are based on the meeting
history of the node. PRoPHET~\cite{lindgren2003probabilistic} assigns a delivery
probability metric to each node which indicates how likely it is that
the message will be delivered to the destination by that particular
node. This metric is updated each time two nodes meet, and thus takes
into account the history of meetings in the network.
MaxProp~\cite{burgess2006maxprop} and MEED~\cite{jones2007practical}
are other examples of history-based algorithms proposed for vehicular
DTNs. In these networks, nodes move with higher speeds, reducing the
amount of time they are in each other's radio range. Hence, the two
main limiting resources are the duration of time that nodes are able
to transfer data and their storage capacities.

Other researchers have examined whether it is possible to exploit
other characteristics of opportunistic networks to improve the
performance of routing algorithms. Since social
interactions often determine when connections between nodes occur,
several algorithms strive to use social network concepts like
betweenness centralities (e.g. \cite{daly2007social}), or community
formations (e.g.
\cite{hui2011bubble,sharma2013contact,xiao2013community}). Other
algorithms have attempted to take advantage of the strategic behaviour
of nodes (e.g. \cite{li2011impact,sermpezis2014understanding}). Our
work focuses on routing a message to a single destination, but there
are connections to research that addresses the task of spreading
information to multiple nodes in a network. Of particular interest is
the gossip-based approach in~\cite{zhang2013gossip}, which greatly reduces the number of
message copies in the network while achieving near-optimal
dissemination.

The experimental-based studies demonstrate the efficiency of their
proposed methods by running simulations on traces recorded from real
world opportunistic networks. Experimental analyses are valuable and
take into account practical considerations, but they can leave us with
an incomplete understanding of how an algorithm operates and how it
will perform in other untested network conditions. For example, the
behaviour of PRoPHET has been shown to be very sensitive to parameter
choice~\cite{grasic2011evolution}. It is also useful to design an
optimal algorithm under slightly less realistic modeling assumptions, and
then consider how it can be adapted to address the practical
limitations, without completely losing its desirable features.
More recent studies have focused on deriving a forwarding process
whose optimality (in some sense) can be mathematically proved under
assumptions about network behaviour.
\cite{conan2008fixed} extends the two hop relay strategy of
\cite{grossglauser2001mobility} by considering the expected delivery
time to the destination as a metric to find the best set of candidate
relays. By increasing the number of relaying steps recursively, a
centralized single-copy multi-hop opportunistic routing scheme is
proposed for sparse DTNs. The main defect of a centralized
approach is that global knowledge of the network is required in
order to make forwarding decisions.

There have been some efforts towards migrating to decentralized
solutions that still provide performance guarantees.
\cite{xiao2013tour} proposes a decentralized time-sensitive algorithm
called TOUR in which message priority is taken into account in
addition to nodes' expected latencies when making forwarding decisions.
Although in TOUR each node only needs to be aware of the local
information about the rates of contacts with its own set of
neighbours, the algorithm assumes that the node knows the exact
contact rates. In most practical scenarios, this assumption is not valid.

Some researchers have explored how imprecision in the measurement
or estimation of network parameters can impact the performance of
opportunistic network routing algorithms. In \cite
{boldrini2010modelling, boldrini2012performance}, Boldrini
et al. discuss different sources of errors that may exist in parameter
estimation like missed encounters, incorrect combination of short
contacts, and memory limitations. They model these errors
as a random variable with a normal distribution and evaluate the
performance of four different forwarding schemes under this
model. Although this error analysis is useful, Boldrini et al. do not
specify how parameters should be estimated in order to obtain a
performance that approaches what can be achieved when perfect a-priori
knowledge of the network parameters is available.

Some of the results in this paper were presented in an earlier
conference paper~\cite{shaghaghian2014opportunistic}, but here we include more extensive experimental
analysis and additional theoretical results.

\section{System Model} \label{sec:system_model}
We consider a network of $N$ mobile nodes which aim to send messages
to a particular destination node $d$. The set of nodes is denoted by
$\mathcal{N}$. We assume that the random inter-meeting times of nodes
are independent and exponentially distributed with parameter
$\lambda_{ij}$ for nodes $i$ and $j$. 

Although the aggregate intermeeting distributions of nodes in mobile
ad-hoc networks often follow a truncated power
law~\cite{cai2009,chaintreau2007impact}, there is evidence that the
intermeeting times of individual pairs of nodes can be adequately
modeled by exponential distributions with heterogeneous
coefficients~\cite{conan2007,gao2009,lee2010,zhu2010}. In particular,
Conan et al.~\cite{conan2007} and Gao et al.~\cite{gao2009} conduct
statistical analyses of mobile social network data traces, including
the Infocom data set~\cite{cambridge-haggle-2006-01-31} that we
analyze in Section~\ref{sec:res}. They demonstrate that most pairs of
nodes have intermeeting times that are approximately exponentially
distributed. In~\cite{lee2010,zhu2010}, approximately exponential
distributions of individual meeting times are detected through
statistical analyses of car/taxi mobility traces.

We associate with the network a {\em contact graph} which is formed by
adding a link between any two nodes that meet. We assume that the
contact graph is connected and denote the set of neighbors of node $i$
in this graph by $\mathcal{S}_i$. Since the contacts between nodes are
not pre-scheduled, we cannot identify end-to-end paths ahead of
time. Hence, solving the routing task is equivalent to identifying the
forwarding decisions that nodes should make when meeting each
other. We assume that nodes' buffer sizes are unlimited, message Time
To Live (TTL) is infinity and that nodes' speed and message lengths
are such that any number of messages can be forwarded during each
meeting.  We consider only algorithms that do not involve
replication. In the class of algorithms we consider, each time node
$i$ meets one of its neighbors $j \in \mathcal{S}_i$, it forwards a
message destined for $d$ with probability $p_{ij}$. Considering the
matrix $\mathbf{P}_{N \times N}$ comprised of all pairs $i$ and $j$,
we set $p_{ij}=0$ if nodes $i$ and $j$ never meet and are thus not
neighbors in the contact graph. We denote the forwarding probabilities
of node $i$ by the vector $\mathbf{p}_i$; this is the $i$-th row of
the matrix $\mathbf{P}$.

The expected latency from node $i$ to destination $d$ is a function of
the probability decision matrix $\mathbf{P}$ and we denote it by
$L_{id}(\mathbf{P})$. Our goal is to find the matrix $\mathbf{P}^*$
such that the sum of the expected latencies of all the nodes in the
network to the specified destination $d$ is minimized. Let us call
this utility function $\mathcal{U}(\mathbf{P}) = \sum_{i\in\mathcal{N}} L_{id}(\mathbf{P})$. We assume that the network topologies and
meeting rates are such that the solution $\mathbf{P}^*$ is unique. If
not, our algorithms guarantee that we reach one of the optimal
matrices, but the proofs are more complicated.
The first step towards achieving this goal and finding matrix $\mathbf{P}^*$ is to discover how the expected latency of an arbitrary node $i$, $L_{id}(\mathbf{P})$, depends on the elements of the probability decision matrix $\mathbf{P}$ in general. Lemma \ref{Lid} provides an expression for $L_{id}(\mathbf{P})$ in
terms of $\mathbf{P}$ and $\lambda_{ij}, j \in \mathcal{S}_i$. The proof is available in Appendix \ref{pl1}.

\begin{lemma} \label{Lid}
The expected latency of a node $i \in \mathcal{N}$ to the destination $d$ is
\begin{equation}\label{lem1}
L_{id}(\mathbf{P})= \frac{1+\sum_{j \in \mathcal{S}_i}p_{ij}\lambda_{ij}L_{jd}(\mathbf{P})}{\sum_{j \in \mathcal{S}_i}p_{ij}\lambda_{ij}}
\end{equation}
\end{lemma}

Based on the expression derived in Lemma \ref{Lid}, the expected latency of each node to the destination depends on the expected latencies of its neighbours. This result raises a substantial question: Does the probability decision  matrix that minimizes the sum of expected latencies of all nodes of the network ($\mathbf{P}^*$), also minimize the expected latency of each individual node? Before continuing to propose algorithms for finding $\mathbf{P}^*$,
we answer this question and make two points about the structure of $\mathbf{P}^*$ through the
following theorem. The proof is provided in Appendix \ref{pt1}.
\begin{theorem}\label{theorem1}
Suppose $\mathbf{P}^*= \arg \min_{\bP \in {[0,1]}^{N\times N}} \sum_{i=1}^N L_{id}(\mathbf{P})$. Then:
\begin{itemize}
\item[(1)] $\mathbf{P}^*$ is a binary matrix (its components are either $0$ or $1$).
\item[(2)] For any $i \in \mathcal{N}$, the matrix $ \mathbf{P}^*$ also
  minimizes $L_{id}(\mathbf{P})$: 
\begin{equation} \label{opt_node}
\forall i \in \mathcal{N}:  \quad   \mathbf{P}^*= \arg \min_{\bP \in {[0,1]}^{N\times N}} L_{id}(\mathbf{P})
\end{equation}
\end{itemize}
\end{theorem}

Theorem \ref{theorem1} shows that the minimization problem is actually a binary problem. Each time node $i$ meets one of its neighbours $j\in \mathcal{S}_i$, it either forwards the message or keeps it. From now on, we change our notation and use the binary indicator matrix $\mathbf{B}$ instead of $\mathbf{P}$ to capture this binary decision. Therefore, the optimization takes the form:
\begin{equation}\label{opt}
\mathbf{B}^*= \arg \min_{\mathbf{B} \in \{0,1\}^{N \times N}} \sum_{i=1}^N L_{id}(\mathbf{B})
\end{equation}
Theorem \ref{theorem1} also states that the optimum solution matrix $\mathbf{B}^*$ can be equivalently achieved by minimizing the expected latency of each of the network nodes to the destination. This is the main idea of developing centralized and decentralized algorithms for finding $\mathbf{B}^*$. In the next section, we introduce the algorithms we have proposed for solving this optimization problem and prove that they find the optimal solution.

\section{Algorithms}\label{sec:algo}

In the first part of this section, we try to to find $\mathbf{B}^*$ in
a centralized fashion where the whole topology and meeting rates of
the network are available at a central unit. This unit calculates a
binary matrix $\mathbf{B}$ and informs the nodes about the neighbours
they should forward their buffered messages to. We prove that the
solution achieved upon completion of this algorithm is the same as the
optimum solution $\mathbf{B}^*$. In the second part of the section,
we introduce a decentralized algorithm and prove that it converges to
the same global solution with
probability $1$. The advantage of the decentralized approach is that
no node needs to have a global knowledge of the network and each node
can learn its own optimal forwarding decisions. The only piece of
information a node needs to know is its meeting rates with its own
neighbours. Finally in the last part of the section, we make our
model more realistic by assuming that nodes have no a-priori knowledge
of any meeting rates. In this more practically realistic scenario,
each node estimates the meeting rates with its neighbours, updating
its estimates each time a contact occurs.

\begin{algorithm}[!htb]
\caption{Centralized Greedy Latency Minimization} \label{algo:cen}
\begin{spacing}{1.0}
\begin{algorithmic}[1]
		\vspace{3mm}
		\STATE{ \texttt{// Initialization}}
		\STATE{$\mathcal{A}= \{d\}$, $\mathbf{B} =
                  \mathbf{0}_{N\times N}$, $L_{dd} = 0$,
                  $L_{jd}=\infty$ for $j \neq d$}
		\STATE{ \texttt{// Iteratively add nodes to the set}}
 		\WHILE{$\mathcal{A} \neq \mathcal{N}$}
			\FOR{each node $i \notin \mathcal{A}$}
				\STATE{Identify $\mathcal{S}_{i\mathcal{A}} =  \mathcal{S}_i \cap \mathcal{A}$}

				\STATE{Calculate $G_{id}$ for  $\mathbf{m}_i^*\in {\{0,1\}}^{|\mathcal{S}_{i\mathcal{A}}|}$\\
				\vspace{0.15cm}
 $G_{id}=\min_{\mathbf{m}_i }\frac{1+\sum_{j \in \mathcal{S}_{i\mathcal{A}}} m_{ij} \lambda_{ij} L_{jd}}{\sum_{j \in \mathcal{S}_{i\mathcal{A}}} m_{ij} \lambda_{ij}}$}
				\vspace{0.15cm}
				\STATE{Denote $\mathcal{D}_i=\{j|j\in\mathcal{S}_{i\mathcal{A}}, m_{ij}^*=1\}$}
			\ENDFOR
 			\STATE{Identify $v=\arg \min_{i \in \mathcal{N/A}} G_{id}$}
			\STATE{Set $L_{vd}=G_{vd}$}
			\STATE{Set $b_{vj}=1$ for all $j \in \mathcal{D}_i$}
			\STATE{Set $\mathcal{A} = \mathcal{A}\cup \{v\}$}
		\ENDWHILE

\end{algorithmic}
\end{spacing}
\end{algorithm}

\subsection{Centralized Approach with Global Knowledge}
Suppose for each node $i \in \mathcal{N}$, the set of neighbours
$\mathcal{S}_i$ and their meeting rates $\lambda_{ij}, j \in
\mathcal{S}_i$ are known at a central calculation unit. Algorithm
\ref{algo:cen} presents an iterative procedure to identify a binary
decision matrix $\mathbf{B}$. In this algorithm, we first decide on
the forwarding rules of the node that has the most frequent direct
contacts with the destination. We refer to this node as node $A$. In
order to achieve the minimum expected latency to the destination, node
$A$ should forward its generated or received messages only to the
destination, ignoring its meetings with any other nodes. All other
nodes that $A$ encounters meet the destination less frequently and, if
they forward their messages to the destination through other nodes,
these other nodes also meet the destination less frequently than
$A$. In subsequent steps of the algorithm, we consider all the nodes
that have direct contacts with the nodes whose forwarding decision
rules have already been made (the set $\mathcal{A}$) and calculate the
minimum latency that each of them can obtain by forwarding through
nodes in $\mathcal{A}$ to the destination. At the end of each
iteration, we finalize the forwarding decision for the one node that
can achieve the minimum latency and add it to $\mathcal{A}$. We repeat
the procedure until the decision has been made for all the nodes of
the network and the elements of the binary matrix $\mathbf{B}$ have
all been specified.  The next theorem states that the binary matrix
$\mathbf{B}$ resulting from applying this procedure, as specified
concretely in Algorithm \ref{algo:cen}, achieves the minimum sum of
expected latencies to the destination. The proof can be found in
Appendix \ref{pt2}.

\begin{theorem}\label{theorem2}
  Suppose all meeting rates are different and there exists
  a unique solution $\mathbf{B}^*$ for the optimization problem \eqref{opt}.
\begin{enumerate}
\item After each iteration of Algorithm \ref{algo:cen},
\begin{enumerate}
\item $\forall i \in \mathcal{A} ,\forall j \in \mathcal{N}: \quad b_{ij}=b_{ij}^*$
\item $\forall i \in \mathcal{A}: \quad L_{id}(\mathbf{B})=L_{id}(\mathbf{B}^*)$
\item $\max_{i \in \mathcal{A}}L_{id}(\mathbf{B}^*)< \min_{i \not\in \mathcal{A}} L_{id}(\mathbf{B}^*) $
\end{enumerate}
\item Upon completion, Algorithm \ref{algo:cen} identifies a labelling $\mathbf{B}$ and associated expected latencies $L_{id}$ such that $\mathbf{B}=\mathbf{B}^*$
\end{enumerate}
\end{theorem}
Theorem \ref{theorem2} demonstrates that the iterative optimization
procedure expressed in Algorithm \ref{algo:cen} finds the solution of
the minimization problem in (\ref{opt}). If there is not a unique
solution, then at some point in Algorithm \ref{algo:cen}, there will
be multiple $\bm_i$ that solve the optimization in line 7. It is
straightforward to show that choosing any one of these $\bm_i$ leads to a
decision matrix $\mathbf{B}$ that achieves the minimum expected latencies. 

\subsection{Decentralized Approach with Partial A-priori Knowledge}\label{dec_partial}
Suppose no central unit exists and each node is just aware of its own
$\mathcal{S}_i$ and the meeting rates $\lambda_{ij}, j\in
\mathcal{S}_i$. Algorithm \ref{algo:DistLatMin} demonstrates how nodes
can make their binary forwarding decisions based on this local
information. Since the expected latency of each node depends on the
expected latency values of its neighbours, nodes need to have an
estimation of their neighbours' expected latencies to be able to make
forwarding decisions. We denote by $\widehat{L}_{id}(j)$ the estimate
at node $j$ of the latency from node $i$ to the destination. In
Algorithm \ref{algo:DistLatMin}, each time two nodes meet, they update
these estimates and then recalculate their optimum forwarding rules. Theorem
\ref{theorem3} proves that this decentralized approach results in the
same global optimum solution. The proof of Theorem \ref{theorem3} is provided in Appendix \ref{pt3}.
\begin{algorithm}[!htb]
\caption{Decentralized Greedy Latency Minimization} \label{algo:DistLatMin}
\begin{spacing}{1.0}
\begin{algorithmic}[1]
		\vspace{3mm}
		\STATE{ \texttt{// Initialization}}
		\STATE{$\mathbf{B} = \mathbf{0}_{N \times N}$}
		\STATE{$ \forall i \in \mathcal{N}/d, \forall j \in \mathcal{N}: \quad  \widehat{L}_{dd}(j)=0$, $\widehat{L}_{id}(j)=\infty$ }
                      \WHILE{ Nodes continue to meet}
	      		\STATE{ \texttt{// Nodes $i$ and $j$ meet at time $t$}}
		           \STATE{ Set $\widehat{L}_{id}(j) = \widehat{L}_{id}(i)$ }
		            \STATE{ Set $\widehat{L}_{jd}(i) = \widehat{L}_{jd}(j)$ }
			\STATE{Update $\widehat{L}_{id}(i)=\min_{\mathbf{m}_i\in
                            {\{0,1\}}^{|\mathcal{S}_i|}}\frac{1+\sum_{k
                              \in \mathcal{S}_i} m_{ik} \lambda_{ik}
                            \widehat{L}_{kd}(i)}{\sum_{k \in \mathcal{S}_i}
                            m_{ik} \lambda_{ik}}$ and identify the minimizing $\mathbf{m}_i^*$ }
			\STATE{Update $\widehat{L}_{jd}(j)=\min_{\mathbf{m}_j\in {\{0,1\}}^{|\mathcal{S}_j|}}\frac{1+\sum_{k \in \mathcal{S}_j} m_{jk} \lambda_{jk} \widehat{L}_{kd}(j)}{\sum_{k \in \mathcal{S}_j} m_{jk} \lambda_{jk}}$ and
                          identify the minimizing  $\mathbf{m}_j^*$ }
			\STATE{Set $\mathbf{b}_i=\mathbf{m}_i^*$ and $\mathbf{b}_j=\mathbf{m}_j^*$}
			\vspace{0.2cm}
		\ENDWHILE
\end{algorithmic}
\end{spacing}
\end{algorithm}

\begin{theorem}\label{theorem3}
The decision matrix $\mathbf{B}$ identified by Algorithm \ref{algo:DistLatMin} converges to $\mathbf{B}^*$ with probability $1$.
\end{theorem}

We refer to our proposed decentralized greedy latency minimization algorithm (Algorithm \ref{algo:DistLatMin}) as MinLat and evaluate its efficiency in different random and real-world networks based on certain performance metrics in Section \ref{sec:res}. Regarding the computational complexity of finding the minimum expected latency in MinLat, the following lemma shows that the optimizations in lines 8 and 9 of this algorithm are linear fractional programs and can be solved quickly using variants from linear programming. Further details are available in Appendix \ref{pl2}. 

\begin{lemma} \label{lem2}
The minimization problem in Algorithm \ref{algo:DistLatMin},
\begin{equation}\label{opt_prob}
\widehat{L}_{id}(i)=\min_{\mathbf{m}_i\in
  {\{0,1\}}^{|\mathcal{S}_i|}}\frac{1+\sum_{k \in \mathcal{S}_i}
  m_{ik} \lambda_{ik} \widehat{L}_{kd}(i)}{\sum_{k \in \mathcal{S}_i}
  m_{ik} \lambda_{ik}} \,\,\,,
\end{equation}
can be converted to a linear programming problem.
\end{lemma}
Assuming that \eqref{opt_prob} can be solved in polynomial order
$P(|S_i|)$, the worst case complexity order of Algorithm
\ref{algo:cen} is $O(N^2)P(N)$ because in the $i$th round of this
algorithm, \eqref{opt_prob} should be solved for each of the $N-i$
nodes that are not in the set $\mathcal{A}$. In Algorithm
\ref{algo:DistLatMin}, each time node $i$ meets one of its
neighbours,it solves a problem of complexity $P(|S_i|)$. The only
information that a node needs to share when it meets another node is
its estimate of its own expected latency to the destination. In the
general case where messages can be destined to any node in the
network, this exchangable message could be a length $N$ vector of
expected latencies to all nodes.

The following proposition provides a
bound on the expected convergence time of Algorithm
\ref{algo:DistLatMin}. The brief proof is provided in
Appendix~\ref{pt3}. The bound depends on the slowest meeting rate
between each node and its candidate relay nodes. This is a
conservative bound, since in practice, a node only needs to meet the
relay nodes to which it actually forwards data under the optimum forwarding rule.
\begin{proposition}\label{prop1}
The expected convergence time, $E(T_N)$, of Algorithm
\ref{algo:DistLatMin} is bounded as 
$
E(T_N)< \frac{1}{\lambda_{1d}} + \sum_{l=2}^{N} \frac{l-1}{\underset{\underset{\lambda_{li}>0}{i
      \in \{1,\dots,l-1\}}}{\min} \lambda_{li}}$.
\end{proposition}

\subsection{Decentralized Approach with No A-priori Knowledge}\label{est}
In part \ref{dec_partial}, we assumed that as soon as a node meets
another node, it has a perfect knowledge of its meeting rate with that
node. In practice, a node will need to estimate its meeting rates with
the neighbours and periodically revise the estimation as meetings
occur (or fail to occur).
Consider an arbitrary pair of nodes that meet each other with rate
$\lambda$. We denote the $i^{\text{th}}$ intermeeting time, which is
the time between $i^{\text{th}}$ and ${i+1}^{\text{th}}$ meetings, by
$x_i \geq 0$. For this specific pair of nodes, $x_i$ is an independent
sample of an exponentially distributed random variable with parameter
$\lambda$. Using the maximum likelihood (ML) approach we can estimate the
parameter $\lambda$ after $n$ samples. The
likelihood function $\mathcal{L}(\lambda|x_1,..,x_n)=\lambda^n e^{-\lambda \sum_{i=1}^n x_i}$ is maximized by 
\begin{equation}
\widehat{\lambda}=\frac{n}{\sum_{i=1}^n x_i}
\end{equation}
Hence, under the exponential model, a node only needs to remember the
last time it met its neighbour and the number of times it has met that
neighbour. With these two pieces of information, it can update its
estimation of the meeting rate ($\widehat{\lambda}^{n}$) from the previously estimated value ($\widehat{\lambda}^{n-1}$) using the following equation.
\begin{equation} \label{estimation}
\widehat{\lambda}^{n}=\frac{n \widehat{\lambda}^{n-1}}{n-1+\widehat{\lambda}^{n-1}x_n}
\end{equation}
Based on this argument, we develop a more practical version of MinLat
in which an arbitrary pair of nodes $i$ and $j$ use their estimated
meeting rates $\widehat{\lambda}_{ij}$ in their calculations and
modify this estimation each time they meet. We refer to this version
of MinLat as MinLat-E.

Let $t$ denote the time since the network began operating, and denote
by $\widetilde{\mathbf{B}}_t$ the decision matrix achieved by MinLat-E
at time $t$. Further, denote by
$\widetilde{L}_{id,t}(\widetilde{\mathbf{B}}_t,i)$ the estimate at
node $i$ at time $t$ of the expected latency to the destination, when
the forwarding decision matrix is $\widetilde{\mathbf{B}}_t$. This
estimate differs from that obtained in Algorithm 2,
$\widehat{L}_{id}(i)$, because the distributed algorithm calculates
them using estimated meeting rates $\widehat{\lambda}_{i,j}$. The
following theorem states that the achieved expected latencies,
$L_{id}(\widetilde{\mathbf{B}}_t)$ and the estimated expected
latencies, $\widetilde{L}_{id,t}(\widetilde{\mathbf{B}}_t,i)$, converge in probability to the
optimum expected latencies $L_{id}(\mathbf{B}^*)$. The proof is
provided in Appendix \ref{pt4}.

\begin{theorem}\label{theorem4}
 For any node in the network, the sequences of estimated and achieved latencies converge to its optimum expected latency in probability, i.e., $\{\widetilde{L}_{id,t}(\widetilde{\mathbf{B}}_t,i)\} \xrightarrow{p} L_{id}(\mathbf{B}^*)$ and $\{L_{id}(\widetilde{\mathbf{B}}_t)\} \xrightarrow{p} L_{id}(\mathbf{B}^*)$. More precisely for any $\epsilon >0$,
\begin{equation}
\underset{t -> \infty}{\lim} P(|\widetilde{L}_{id,t}(\widetilde{\mathbf{B}}_t,i)-L_{id}(\mathbf{B}^*)|< \epsilon)=1
\end{equation}
\begin{equation}
\underset{t -> \infty}{\lim} P(|L_{id}(\widetilde{\mathbf{B}}_t)-L_{id}(\mathbf{B}^*)|< \epsilon)=1
\end{equation}
\end{theorem}
 We check the claims of Theorem \ref{theorem4} and investigate the
 convergence speed of MinLat-E through simulations in Section
\ref{sec:res}. 

\section{Simulation Results}\label{sec:res} 
In this section, we investigate the efficiency of our
proposed approach in modeling and solving the forwarding/routing problem in
different opportunistic network scenarios. 
We first test our algorithms using three different
networks to model the contacts between $N=41$ mobile nodes.
The characteristics of the networks are derived from the Infocom05 dataset~\cite{cambridge-haggle-2006-01-31}. This data set is based on an
experiment conducted during the IEEE Infocom 2005 conference in Miami
where 41 Bluetooth enabled devices (Intel iMote) were carried by
attendees for 3 days. The start and end times of each contact
between participants were recorded.  The average time between node
contacts in the Infocom05 dataset is $1.3 \times 10^4$ seconds ($3.7$
hours). In our processing, we only consider the contacts in which both devices
recognized each other so that an acknowledged message could be
transfered between them. 

In the first network, ({\em Net I}), we construct a contact graph using
an evolving undirected network model based on the preferential
attachment mechanism. We start with a small fully connected graph of
$m_0 = 5$ vertices and add vertices to it one by one until the
graph consists of $N=41$ nodes. At each step, the new vertex is
connected to $m=5$ previously existing vertices. The probability
that the new vertex is connected to vertex $i$ is $\frac{k_i}{\sum_j
  k_j}$ where $k_i$ is the degree of $i$ up to this stage. After
building the contact graph, we assign a parameter $\lambda_{ij}$ to
each pair of nodes $i$ and $j$ which are connected in the contact
graph and assume that they meet with exponentially distributed
intermeeting times with parameter $\lambda_{ij }$. We choose the parameters $\lambda_{ij}$ from a uniform distribution with the same expectation as the average of node meeting rates observed in the Infocom05 dataset. 

In {\em Net II}, we set $\lambda_{ij}$ to be equal to the inverse of
the average intermeeting time between nodes $i$ and $j$ in the
Infocom05 dataset. We are interested in the behaviour of the
algorithms in relatively sparse networks, so we limit the number of
neighbours of each node: node $i$ is only connected to node $j$ in the
contact graph if the meeting rate $\lambda_{ij}$ is among the largest
$K=10$ meeting rates of either node $i$ or node $j$. In our simulations, the meeting
times between nodes $i$ and $j$ for {\em Net II} are then chosen from an exponential
distribution with parameter $\lambda_{ij}$. In the third experimental
network, {\em Net III}, we use the actual meeting times recorded in
the Infocom05 dataset. The analysis in~\cite{gao2009} indicates that
the distribution of individual intermeeting times for most pairs of
nodes can be approximated reasonably well by an exponential
distribution; on the other hand, the aggregate distribution of contact
times shows heavy-tailed behaviour and is better approximated using a
truncated power distribution~\cite{chaintreau2007impact, cai2009}.
Table \ref{table:networks} summarizes the properties of the test networks.

\begin{table}[hbt] 
\small
\caption{Test Network Properties} 
\centering 
\begin{tabular}{|c|c|c|c|c|} 
\hline
 \begin{tabular}{@{}c@{}}Net- \\work\end{tabular} & \begin{tabular}{@{}c@{}}Contact \\Graph\end{tabular} &  Parameters & \begin{tabular}{@{}c@{}}Intermeeting \\Times\end{tabular}& \begin{tabular}{@{}c@{}}Number \\of nodes\end{tabular}\\ [0.5ex] 
\hline 
{\em I} & \begin{tabular}{@{}c@{}} Preferential  \\Attachment\end{tabular} & \begin{tabular}{@{}c@{}} $m_0=5$  \\$m=5$ \end{tabular} & \begin{tabular}{@{}c@{}} Exponential  \\ $\lambda$: uniform\end{tabular} & $41$ \\[0.3ex]
\hline
{\em II} & \begin{tabular}{@{}c@{}} Sparsified  \\Infocom \end{tabular} &$K=10$& \begin{tabular}{@{}c@{}} Exponential  \\$\lambda$: data-set\end{tabular} & $41$ \\[0.3ex]
\hline
{\em III} & \begin{tabular}{@{}c@{}} Sparsified  \\Infocom \end{tabular} &$K=10$& Data-set times & $41$ \\[0.3ex]
\hline
\end{tabular} 
\label{table:networks} 
\end{table} 

As mentioned in Section \ref{sec:algo}, we call our proposed
decentralized greedy latency minimization algorithm {\em MinLat} and refer
to its more practical version with meeting rate estimations as
{\em MinLat-E}. In these two algorithms, the decisions that nodes make for
future forwarding rules depend on the (estimated) meeting rates, which
are derived from the frequency of past contacts between nodes. Thus, MinLat and
MinLat-E can be identified as history-based routing algorithms. In
order to evaluate their performance, we compare them with
existing history-based routing protocols that can be
implemented in a distributed fashion and do not need a-priori
knowledge of the network topology.  As mentioned in Section
\ref{sec:intro}, the fixed point algorithm proposed in
\cite{conan2008fixed} is proved to result in the minimum expected
latency (which is expected to be the same as the result of our
proposed centralized Algorithm \ref{algo:cen}). However, the proposed
algorithm in \cite{conan2008fixed} is centralized and needs to be
performed in a control unit where the whole topology of the network is
known. The TOUR algorithm proposed in \cite{xiao2013tour} is
decentralized, but each node needs perfect a-priori
knowledge of its meeting rates with other nodes. Also, the main focus
of TOUR is to find the optimum way to make forwarding decisions
based on the priorities of messages. We have chosen PRoPHETv2 \cite{grasic2011evolution} and MaxProp
\cite{burgess2006maxprop} as the most appropriate candidates for
comparison. We also compare to Epidemic
routing~\cite{vahdat2000epidemic}, which is expected
to result in a high delivery probability at the expense of high usage
of network resources. The parameters of PRoPHETv2 are set to those
suggested in \cite{lindgren2003probabilistic} and
\cite{grasic2011evolution}, i.e., $P_{init}=0.75$, $\beta=0.25$,
$\gamma=0.98$, and $\text{time step}=1$. In order to put the focus on
evaluation of the performance efficiency of forwarding rules and
eliminate the effect of the buffer management technique on this
performance, we first test the algorithms on ideal network scenarios
where the message life times, buffer sizes and data exchanges have no
restriction. Therefore for these simulations, the dropping rules proposed
for MaxProp in \cite{burgess2006maxprop} are not applied. We then
study the network behaviour when these practical challenges are added
to the simulation setups.

\begin{figure}[!htb]
\centering
\subfloat[Average Latency (sec)]{\includegraphics[width=1.69in,height=1.7in]{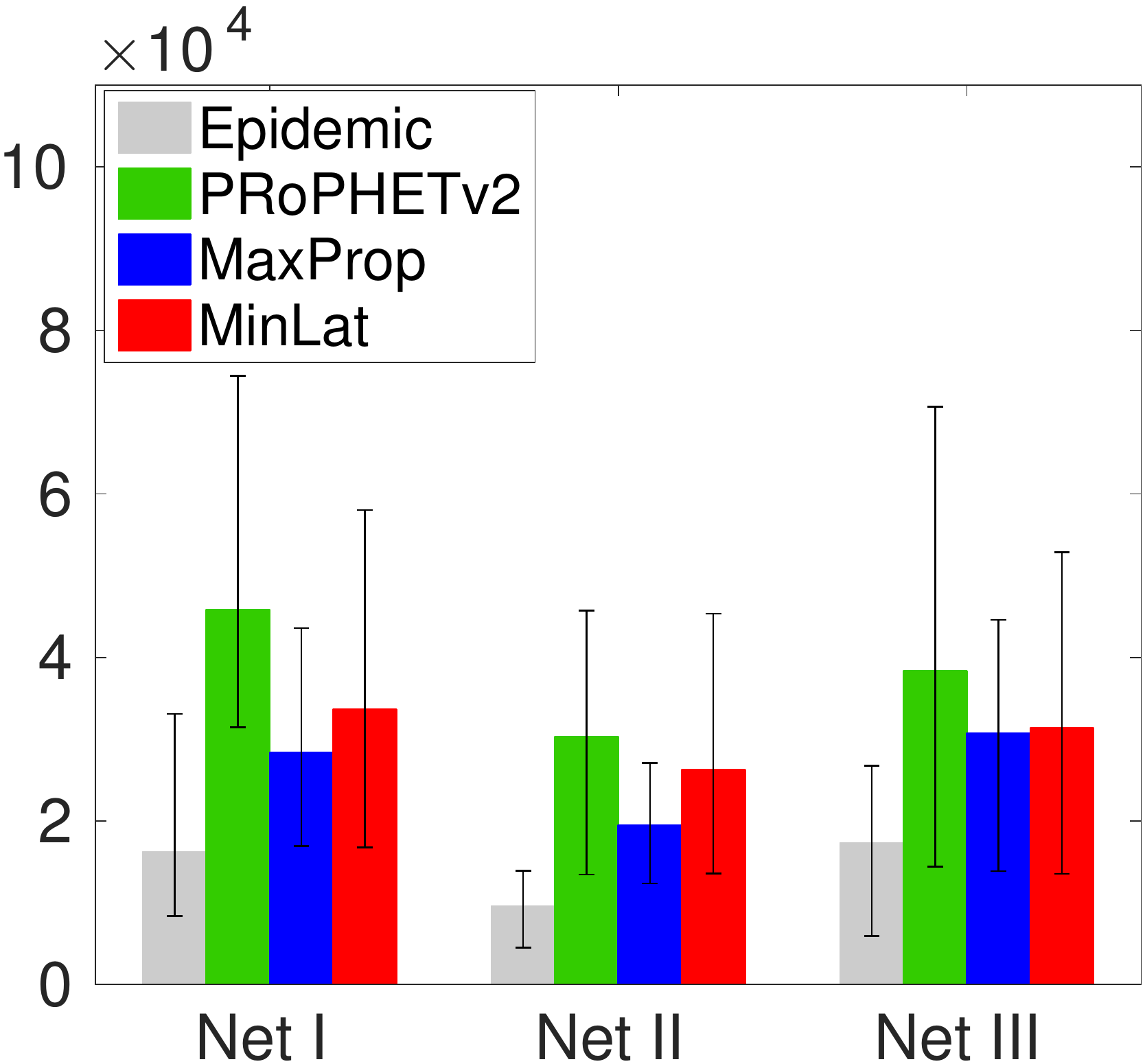}\label{fig:fig1_first_case}}\hspace*{0.08in}
\subfloat[Average Delivery Rate]{\includegraphics[width=1.68in,height=1.61in]{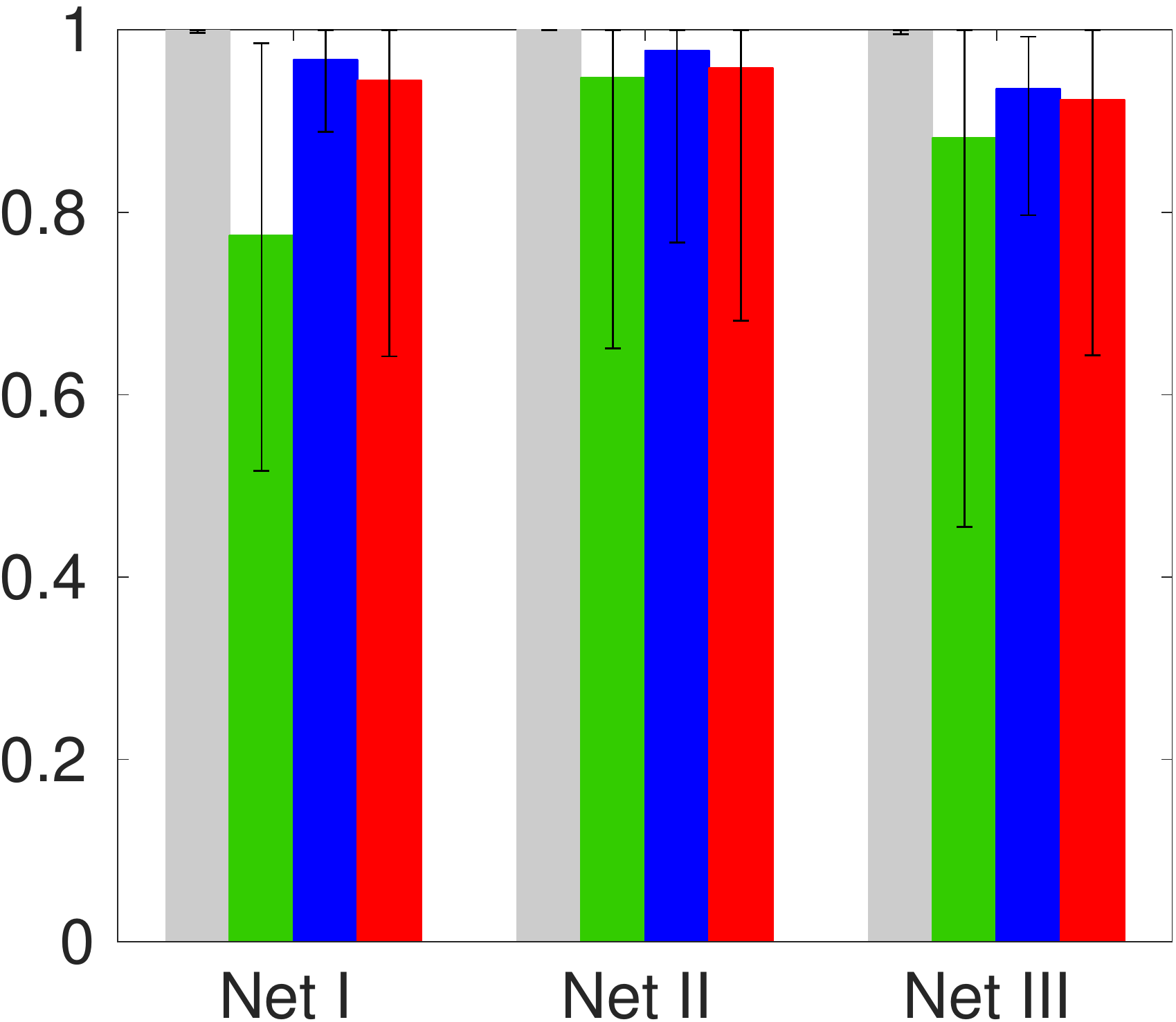}\label{fig:fig1_second_case}}\\

\hspace*{0.009in} \subfloat[Average Hop Count]{\includegraphics[width=1.65in,height=1.62in]{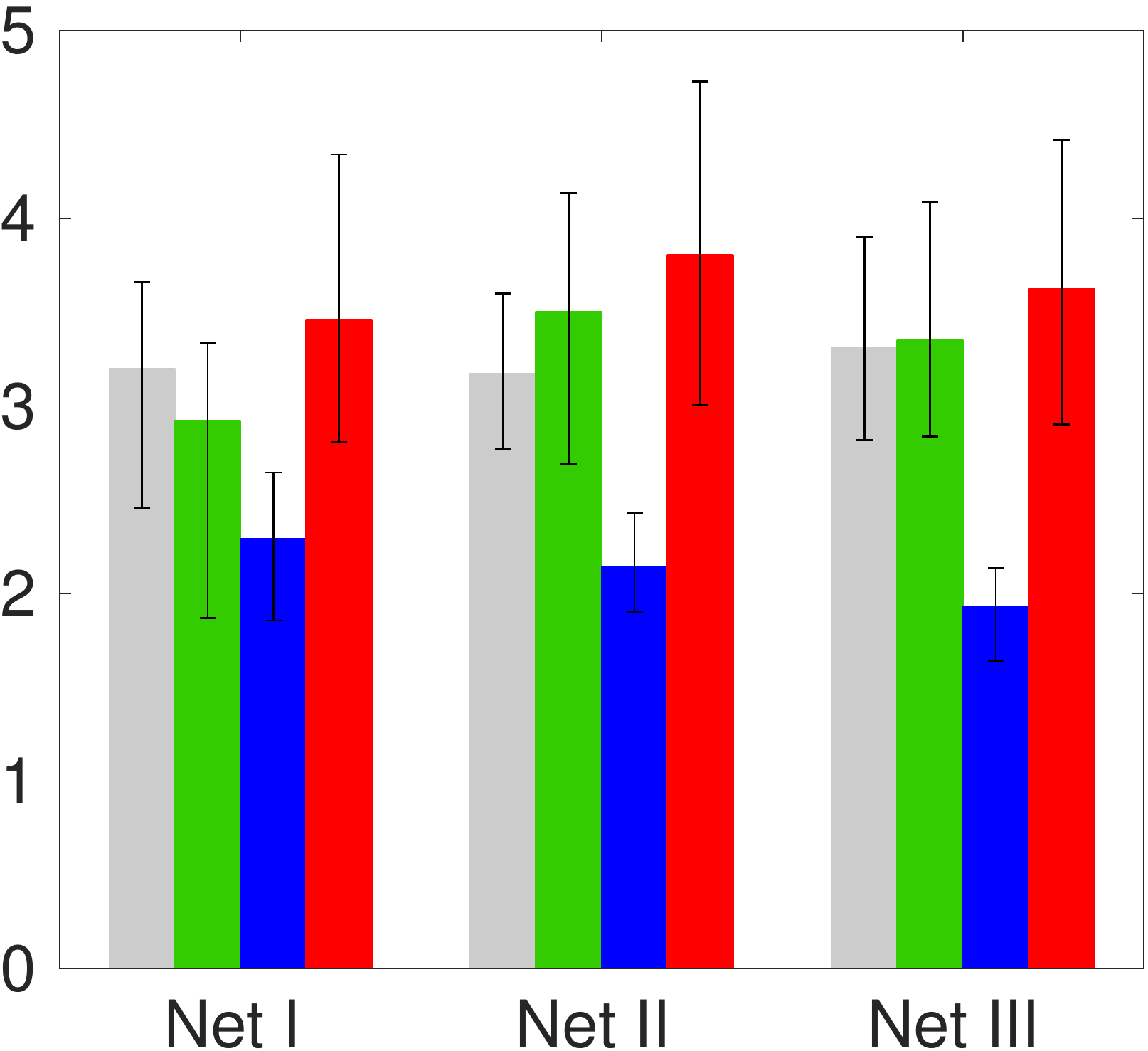}\label{fig:fig1_third_case}}\hspace*{0.05in}
\subfloat[Average Buffer Occupancy]{\includegraphics[width=1.72in,height=1.6in]{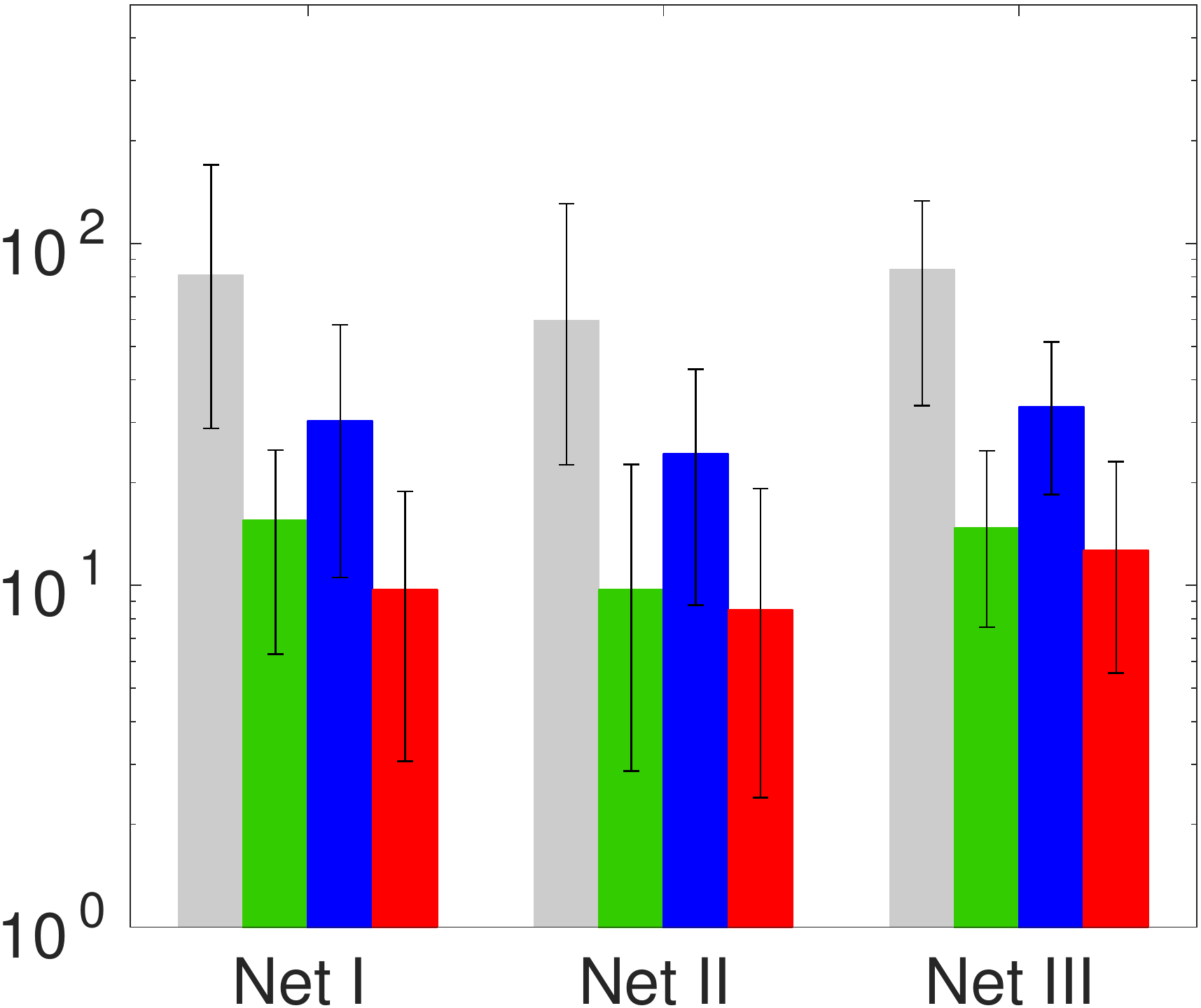}\label{fig:fig1_fourth_case}}
\caption{Comparison Metrics in Test Networks with $N=41$}
\label{fig:res1}
\end{figure}

We divide the Infocom05 dataset into slots of 12 hours. In each of
these time slots, we build networks \textit{I} to \textit{III} using
the nodes that are present in that period. The intermeeting time
exponential parameters ($\lambda_{ij}$s) are estimated based on the
meetings that occurred in that specific time slot and networks
\textit{I} and \textit{II} are constructed using these estimated
parameters. For each of the first four 12-hour periods (the first two
days of the conference), we send
$1000$ messages, spaced by $5$-second intervals, from randomly chosen
source nodes to a particular destination. We terminate the simulation
at the end of the 3-day period, and calculate the fraction of messages
successfully delivered by each of the single copy (PRoPHETv2, MinLat)
and multi-copy (Epidemic, MaxProp) forwarding algorithms. For each
algorithm, we also calculate the average latency of the messages that
are delivered by all four algorithms; the average number of hops that
messages pass to reach the destination; and the average buffer
occupancy of the nodes. For each of the $41$ nodes of the network, we calculate the average of
performance metrics over the time slots when that node has been chosen
as a destination. Figure \ref{fig:res1} shows the average and the
$95\%$ confidence intervals of the four performance metrics for
different destinations in the three test networks using Epidemic,
PRoPHETv2, MaxProp, and MinLat forwarding algorithms.
There is no restriction on message life time or buffer size that can
cause message dropping in these simulations. However, the delivery
rates in some cases are less than $1$ because the simulations are
terminated before all of the generated messages are successfully
delivered. We observe that in all the three test networks, MinLat has
a better performance than the other existing history-base single copy
algorithm, PRoPHETv2, in terms of delivery rate and average
latency. Its performance is also comparable to MaxProp which is a
history-based multi-copy algorithm. Also, noting that the scale of the
vertical axis of Figure \ref{fig:fig1_fourth_case} is logarithmic, we
see that MinLat occupies much less memory of nodes' buffers in
average. For networks {\em I} and {\em II}, where the assumption of
exponentially distributed intermeeting times holds, delivering a
higher rate of messages with lower average latencies than PRoPHETv2 is
expected from Theorem \ref{theorem3}. However, we observe that this
result also holds for network {\em III} where the actual meeting times
are used. All algorithms display slightly poorer performance in
network {\em III}; this is probably due to heavy-tailed and
non-stationary intermeeting times.

\begin{figure}[!hbt]
\centering
\includegraphics[width=3in]{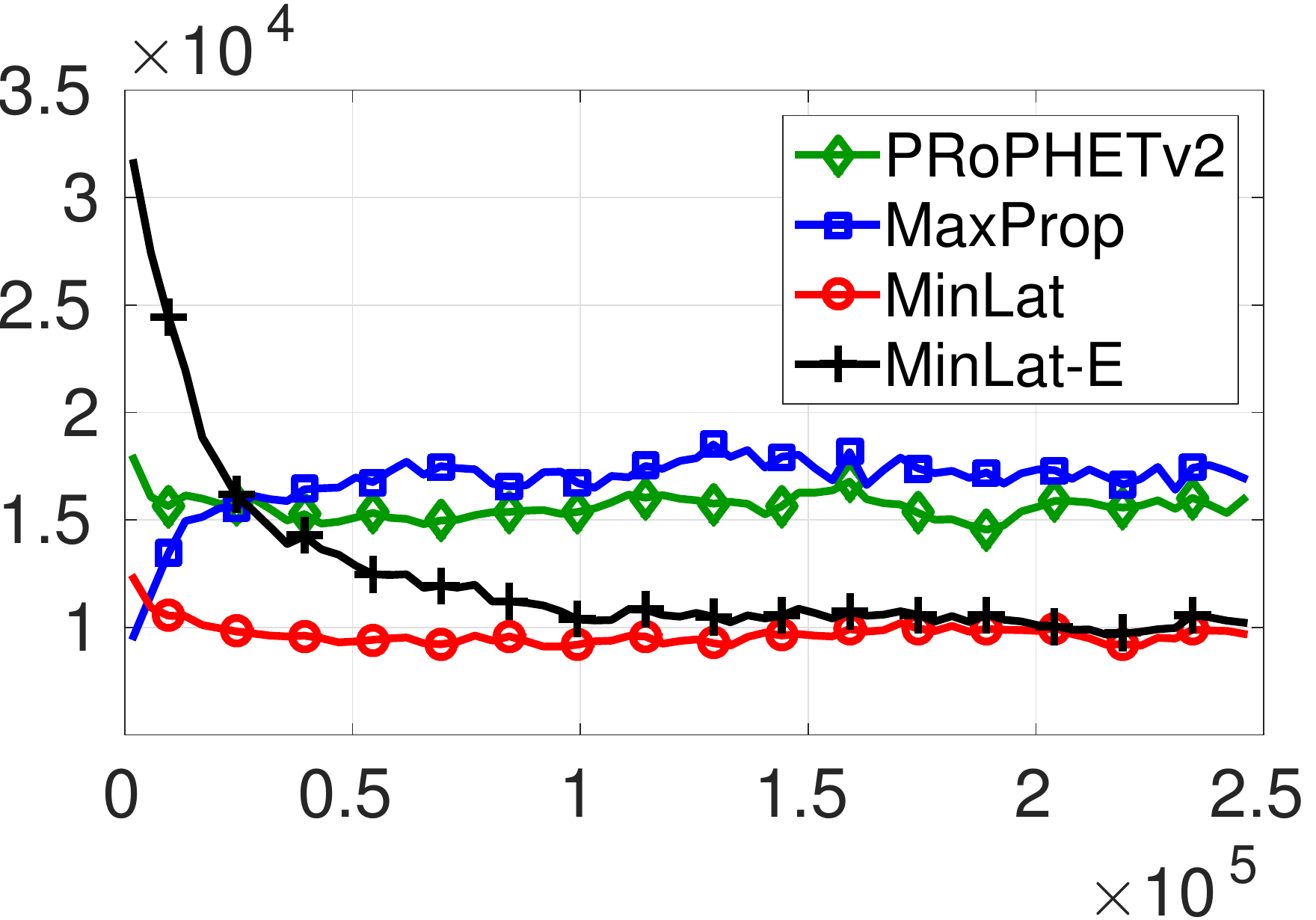}
\caption{Evolution of average delivery latency with time (both in seconds).}
\label{fig:fig2}
\end{figure}

In order to explore how the incorporation of meeting rate estimations
in MinLat-E affects the message delivery performance, we conduct
further simulations with a different message generation scenario on
network {\em II}. Figure \ref{fig:fig2} displays the average delivery
latency as a function of time for the history-based routing algorithms
PRoPHET, MaxProp, MinLat, and MinLat-E. The delivery latency values are
averaged over $32$ simulation runs while the destination node and
message generation times are fixed in all the runs. Messages are
generated with interarrival time of $t$ seconds at randomly chosen
source nodes where $t$ is uniformly distributed in $[0,5]$. Each point
on the curves represents the average latency of the $1500$ most
recently sent messages. As Figure \ref{fig:fig2} shows, in all of the
three single copy routing algorithms (PRoPHET, MinLat, MinLat-E), the
average time it takes for a message to be delivered at the destination
decreases as time goes by. This decreasing trend is due to the fact
that the forwarding rules discovered by nodes improve as the
nodes have more contacts and their information concerning their neighours'
message delivery capabilities becomes more accurate. However, in the
multi-copy routing algorithm, MaxProp, the average message delivery
time increases in the beginning. In MaxProp, the weights assigned to the
links are initialized to be equal, which means that nodes forward
messages to more of their neighbours. There is thus a high level of
message replication which leads to messages reaching the destination
sooner on average. As time passes, the level of replication decreases and the
average delivery time increases. We also observe that MinLat-E and MinLat
eventually achieve the same average delivery latencies, as expected from Theorem
\ref{theorem4}. However, the convergence to the optimum point is
slower in MinLat-E due to the time it takes for nodes to obtain
accurate estimates of their meeting rates.

\begin{figure}[!hbt]
\centering
\hspace*{0.02in} \subfloat[Average Latency (sec)]{\includegraphics[width=1.64in,height=1.7in]{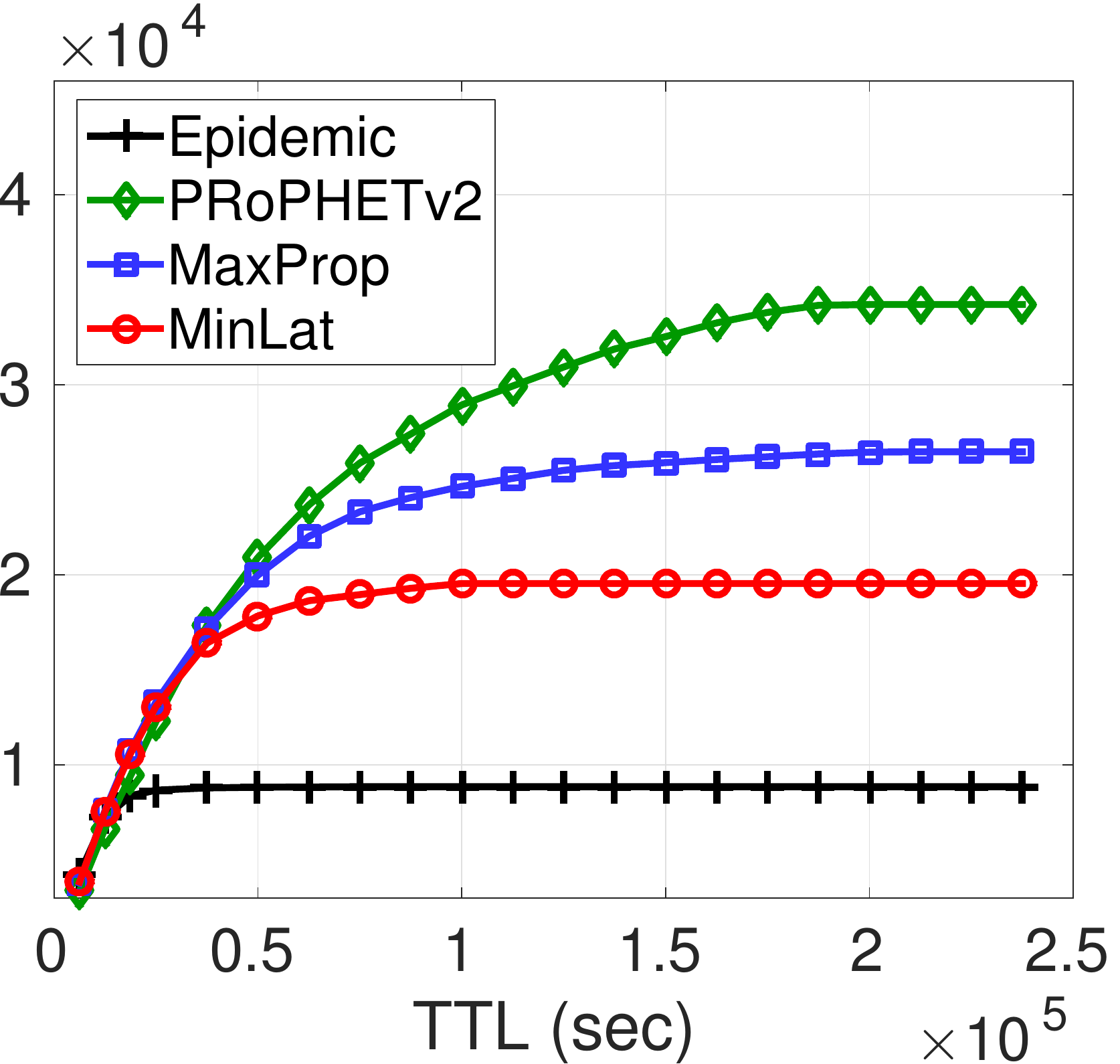}\label{fig:TTL_first}}\hspace*{0in}
\subfloat[Average Delivery Rate]{\includegraphics[width=1.7in,height=1.62in]{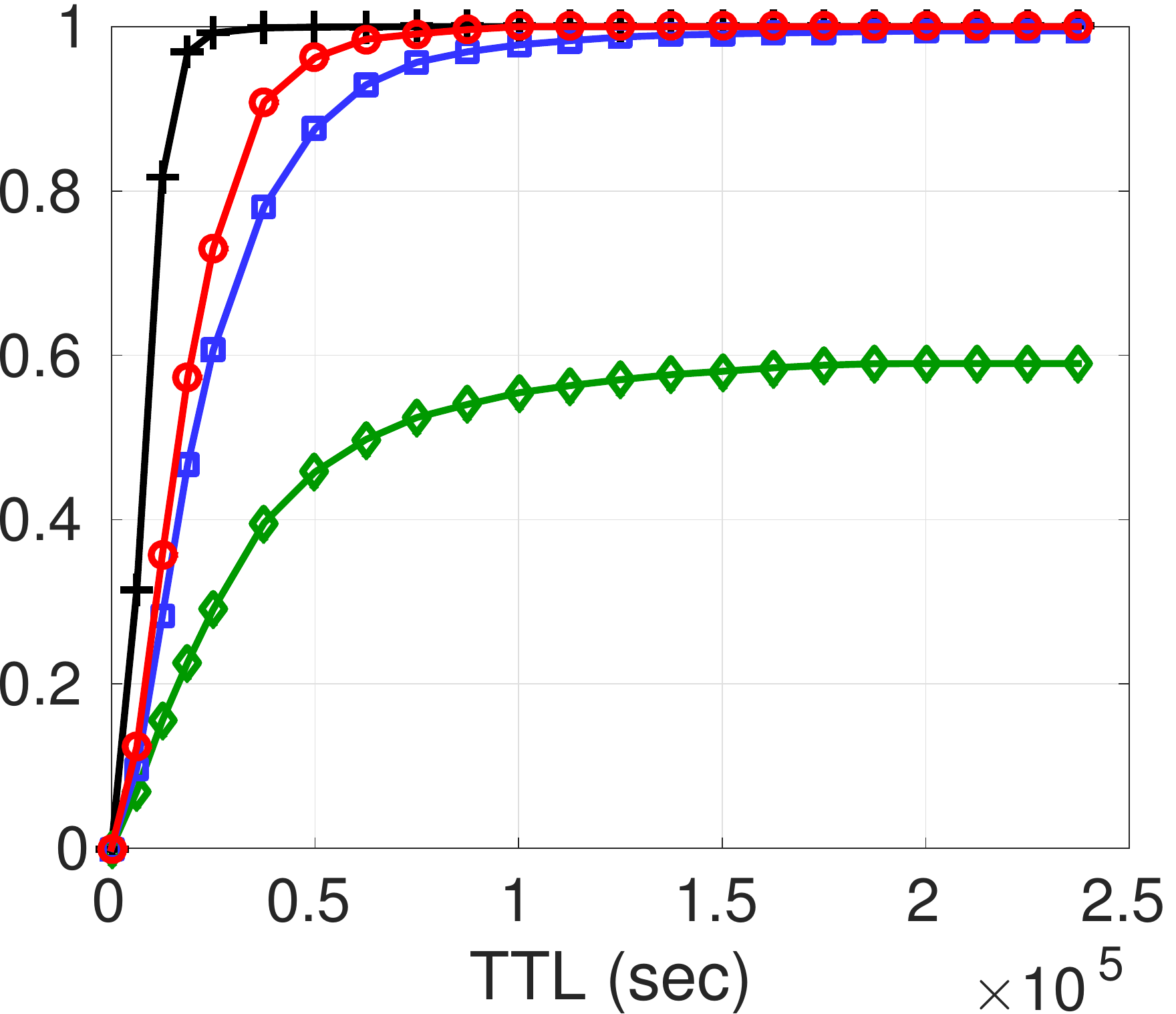}\label{fig:TTL_second}}\\

\subfloat[Average Hop Count]{\includegraphics[width=1.7in,height=1.6in]{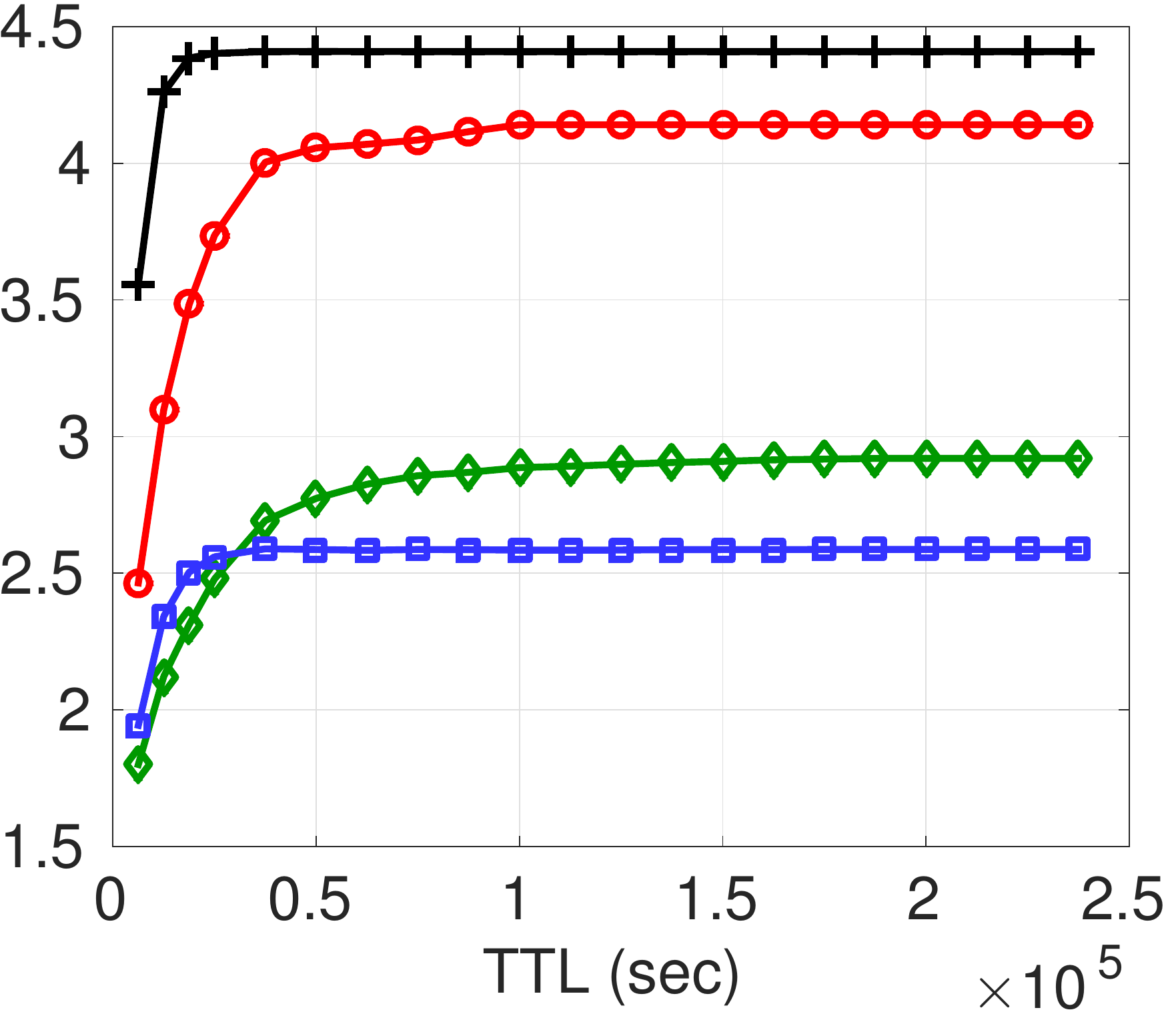}\label{fig:TTL_third}}\hspace*{0in}
\subfloat[Average Buffer Occupancy]{\includegraphics[width=1.7in,height=1.6in]{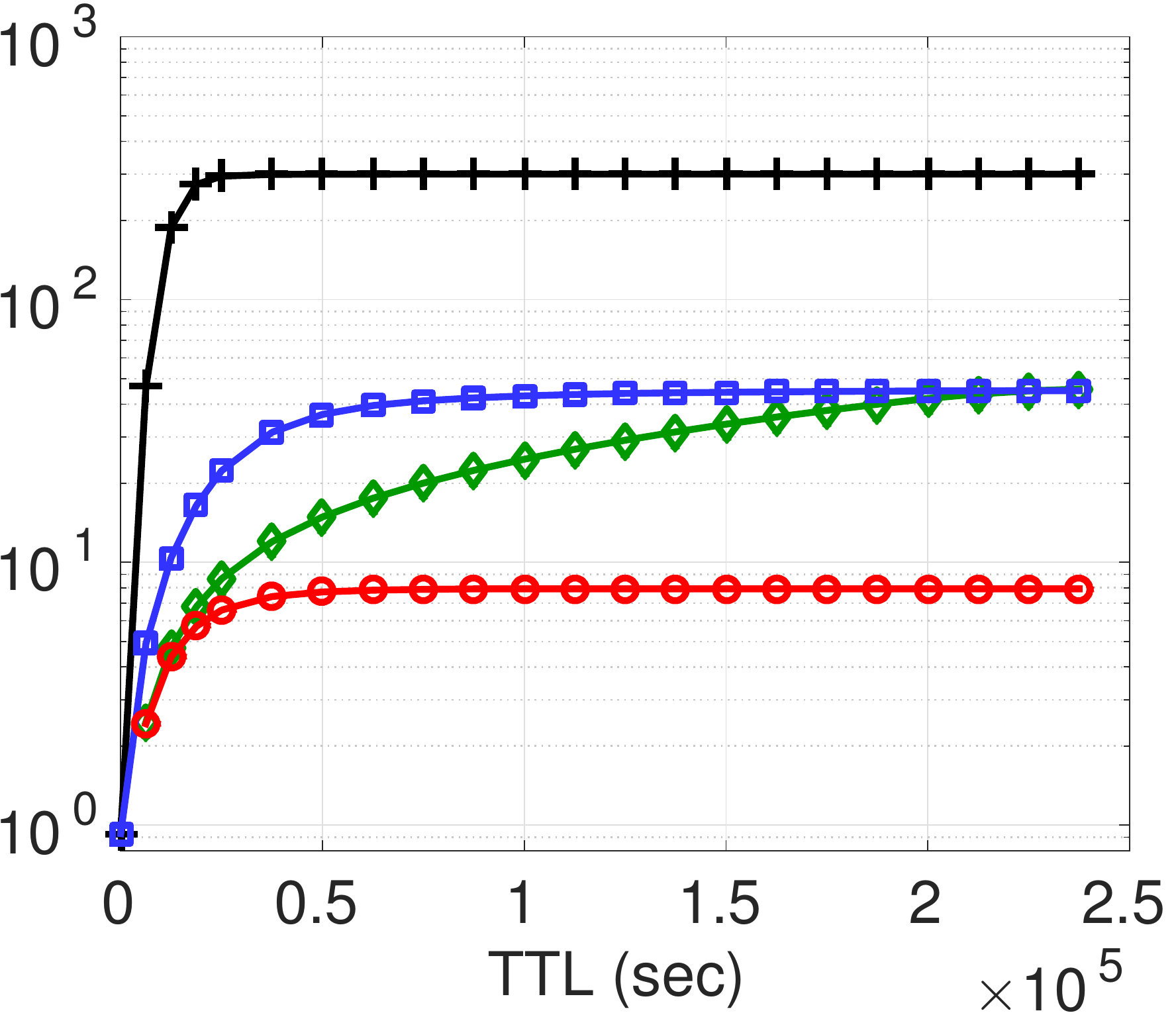}\label{fig:TTL_fourth}}
\caption{Effect of TTL on Performance Metrics}
\label{fig:res_TTL}
\end{figure}

We examine the performance of the forwarding algorithms in larger
networks by extending network {\em I} to $100$ nodes but with the same
average $\lambda$ parameter for exponential intermeeting times. We
also make our model more realistic by adding some practical
restrictions to the network model. First, we assume that messages have
finite TTL, i.e., a message is discarded when its lifetime exceeds a
certain threshold. Figure \ref{fig:res_TTL} displays the performance
of routing algorithms for different values of TTL varying from $0$ to
as large as the simulation time (almost $2.5 \times
10^5$seconds). Simulations are run $100$ times and in each round, a
different destination is randomly chosen from network nodes based on a
uniform distribution.  The average latency and average hop count are
calculated only for the messages that reach the
destination. 

The simulation results in Figure \ref{fig:TTL_second} show that
decreasing the TTL has a similar overall effect on all of the
algorithms. For larger TTLs, the delivery rate increases, but the
buffer occupancy, average latency and hop count also increase. MinLat
outperforms PRoPHET and MaxProp in terms of delivery rate, average
latency, and buffer occupancy even in a scenario with a restricted
message life time.  Although intermeeting times are exponentially distributed and the
contact graph is based on preferential attachment, in this larger
network of 100 nodes, PRoPHET cannot reach $100$ percent delivery rate
even without any restriction on TTL.

\begin{figure}[hbt]
\centering
\hspace*{0.03in} \subfloat[Average Latency (sec)]{\includegraphics[width=1.6in,height=1.7in]{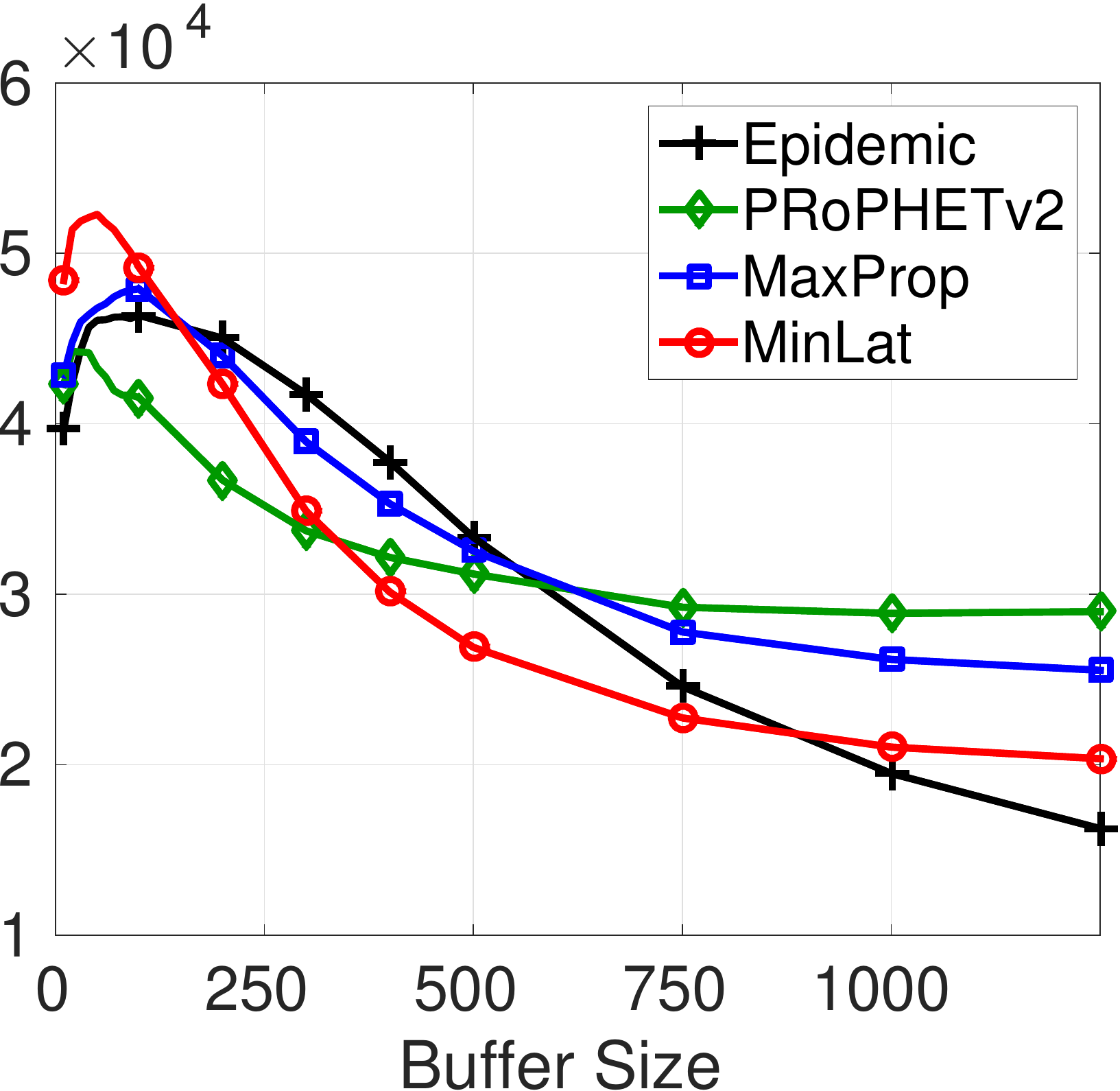}\label{fig:buff_first}}\hspace*{0 in}
\subfloat[Average Delivery Rate]{\includegraphics[width=1.7in,height=1.62in]{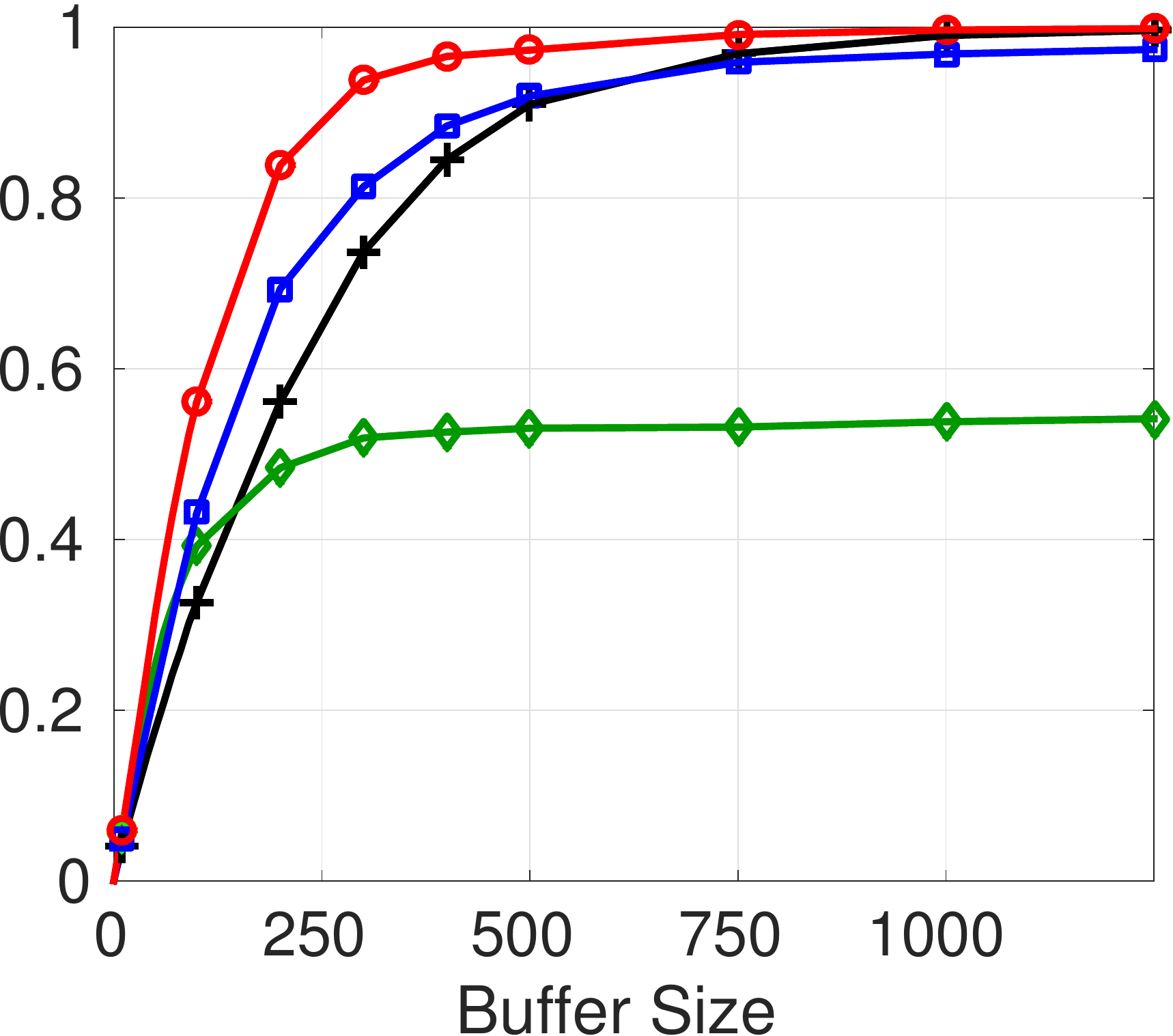}\label{fig:buff_second}}\\
\subfloat[Average Hop Count]{\includegraphics[width=1.67in,height=1.6in]{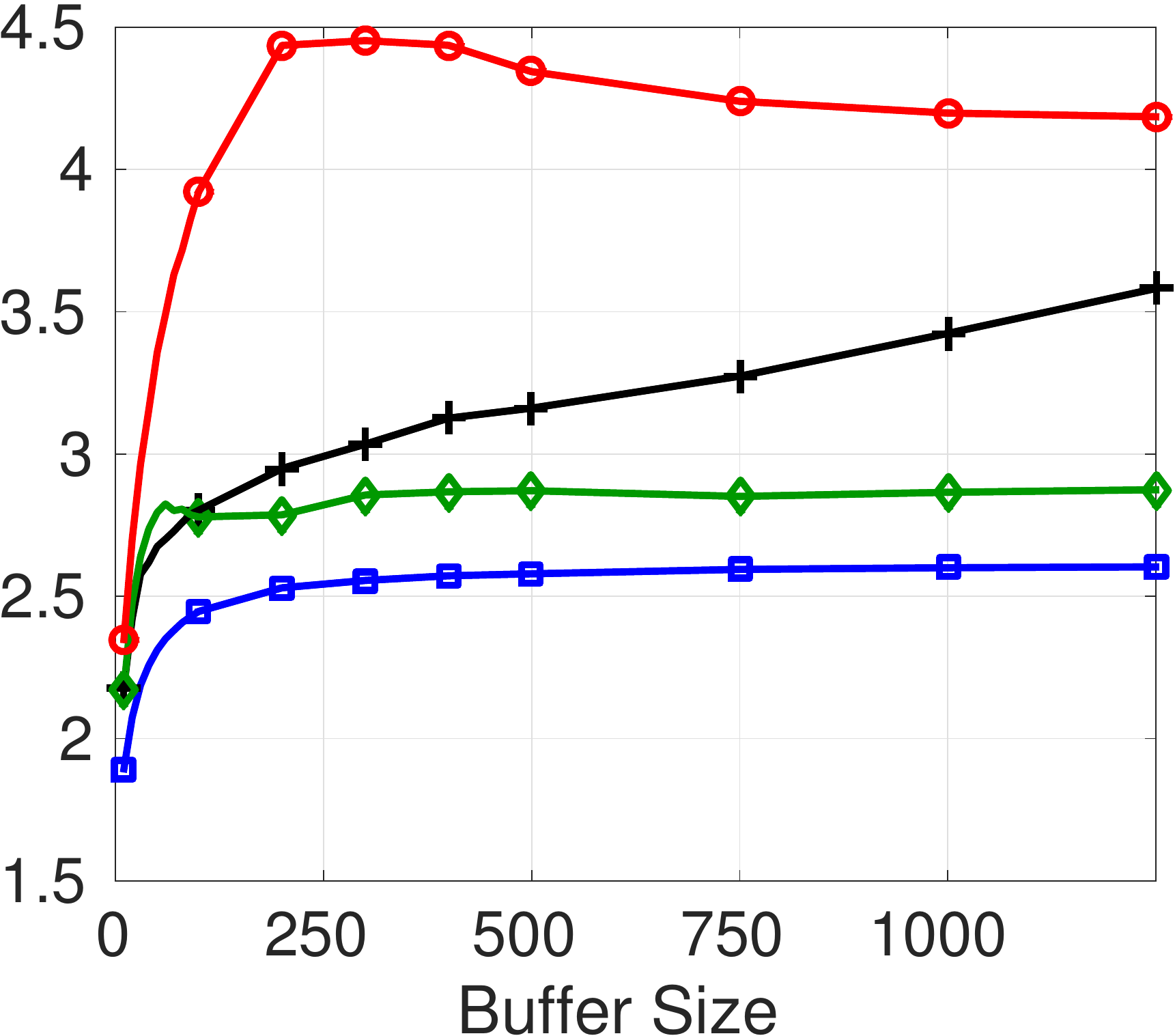}\label{fig:buff_third}}\hspace*{0in}
\subfloat[Average Buffer Occupancy]{\includegraphics[width=1.7in,height=1.62in]{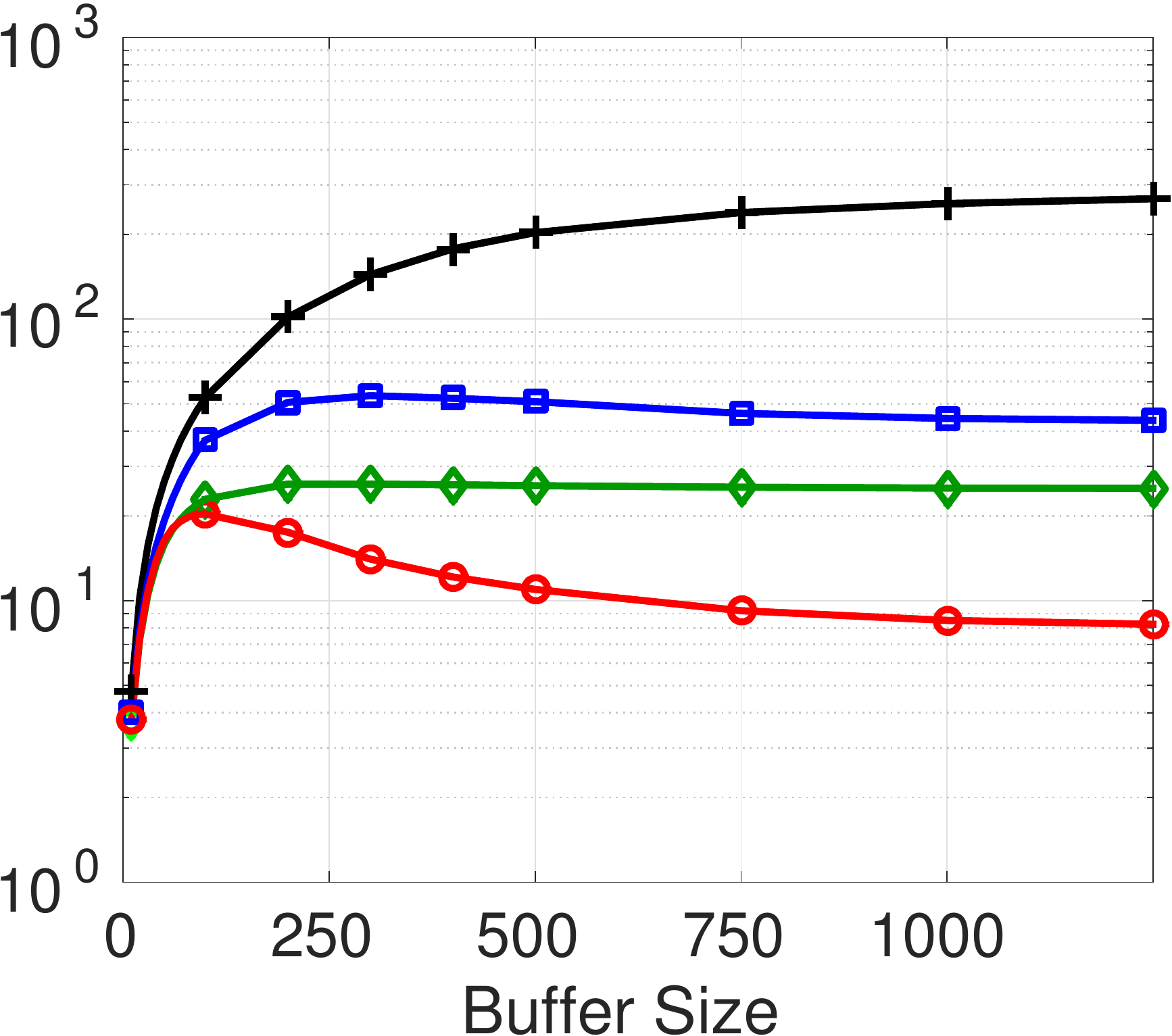}\label{fig:buff_fourth}}
\caption{Effect of Buffer Size on Performance Metrics}
\label{fig:res_buff}
\end{figure}

The next step towards a more realistic network model is to consider
limits on buffer size. In the next set of simulations, we assume that TTL is $10^5$
seconds so that all algorithms reach their best possible delivery
rate. We also assume that each node has a limited capacity for keeping
the messages. When the buffer occupancy of a node reaches its limit,
messages from other nodes are not forwarded to it. Moreover, any generated
messages at the fully occupied node are immediately dropped. Figure
\ref{fig:res_buff} shows the performance of algorithms for buffer
sizes in the range of $0$ to $1250$ messages. We see that increasing
the buffer size improves the delivery rate for all algorithms. It also
reduces the average latency because as the buffer
size increases, nodes can more frequently follow their optimum
forwarding rules.

\begin{figure}[!htb]
\centering
\hspace*{0.03in}\subfloat[Average Latency (sec)]{\includegraphics[width=1.6in,height=1.7in]{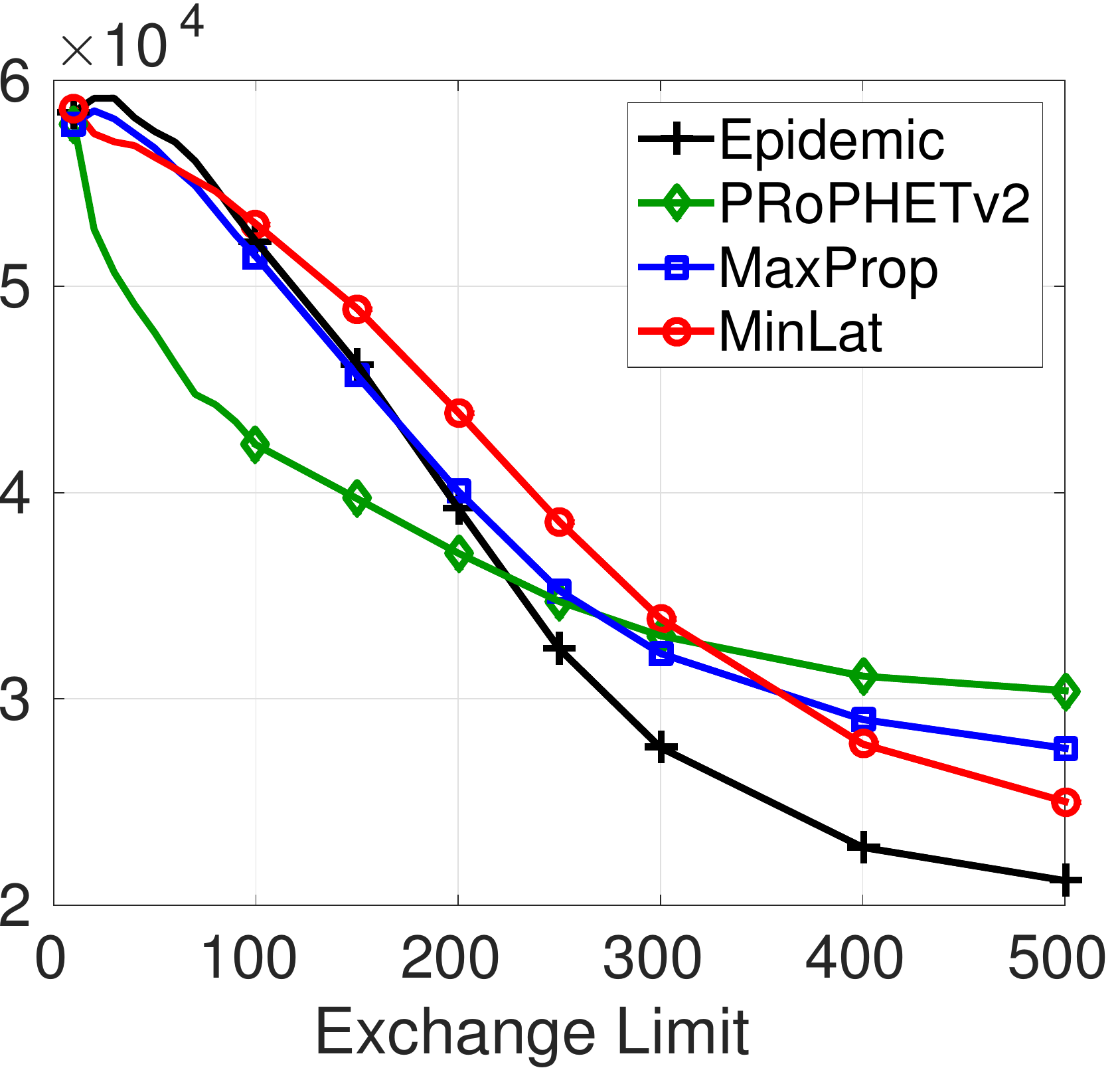}\label{fig:exchange_first}}\hspace*{0 in}
\subfloat[Average Delivery Rate]{\includegraphics[width=1.7in,height=1.6in]{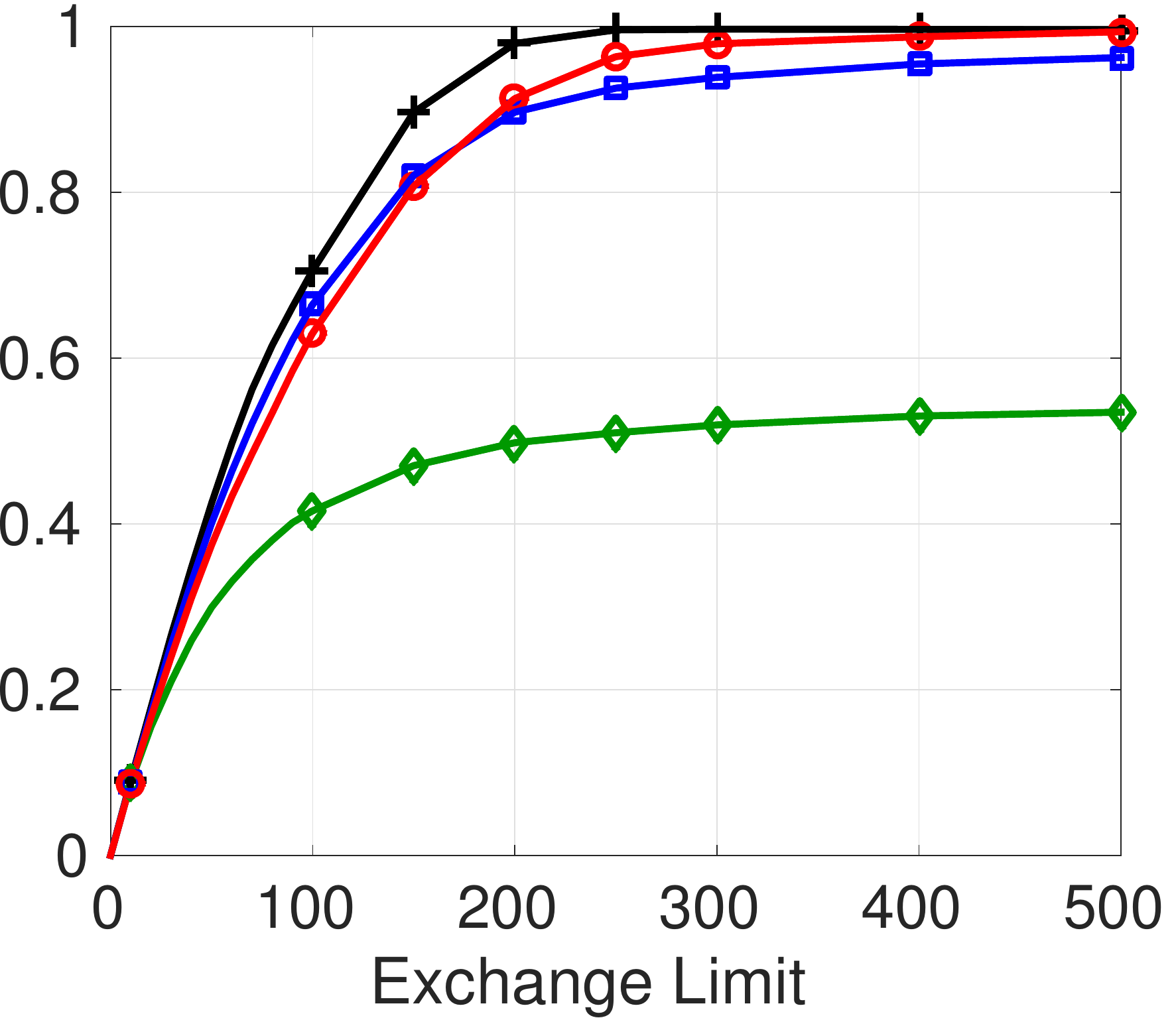}\label{fig:exchange_second}}\\
\hspace*{0.03in}\subfloat[Average Hop Count]{\includegraphics[width=1.6in,height=1.6in]{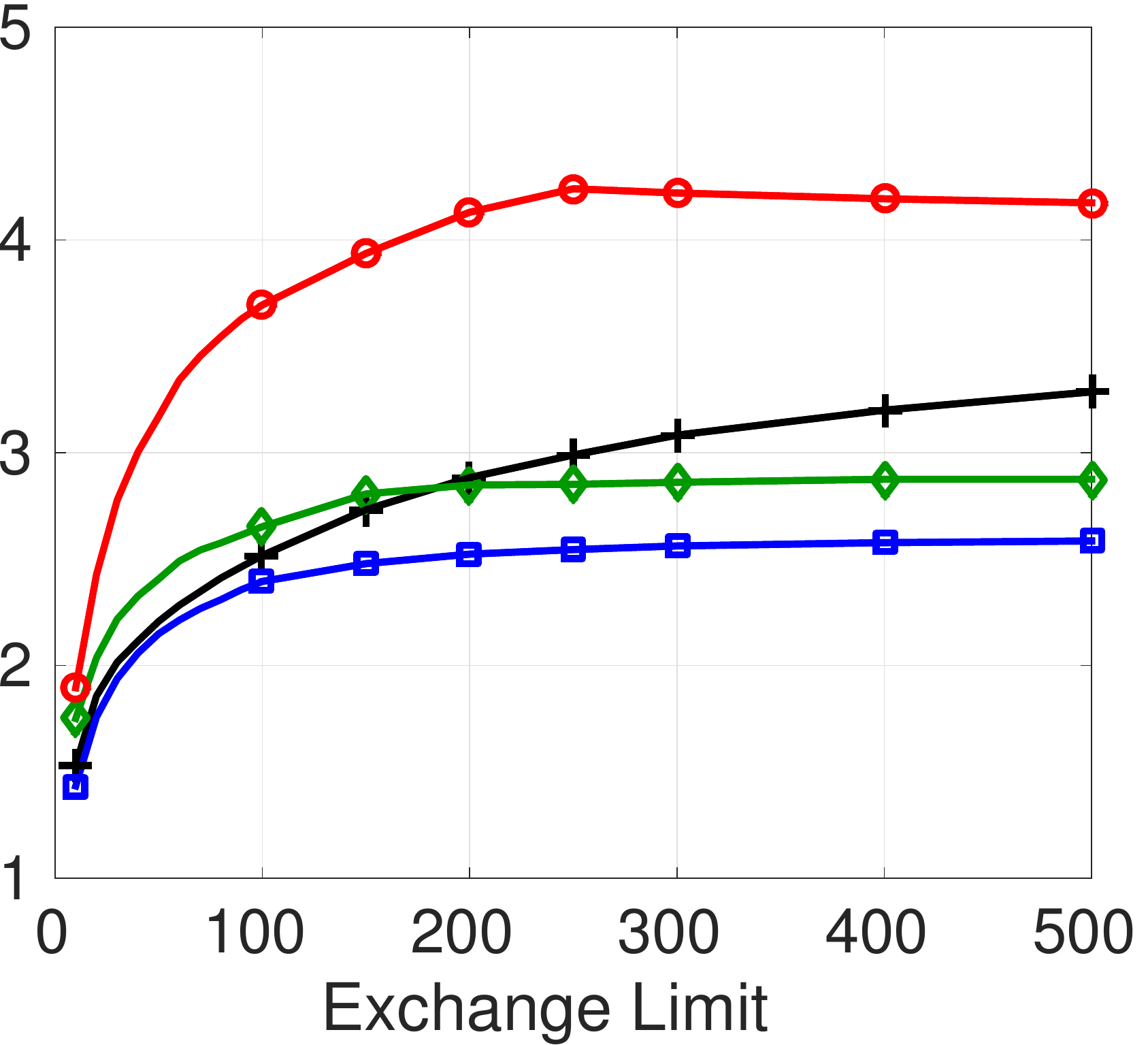}\label{fig:exchange_third}}\hspace*{0in}
\subfloat[Average Buffer Occupancy]{\includegraphics[width=1.7in,height=1.62in]{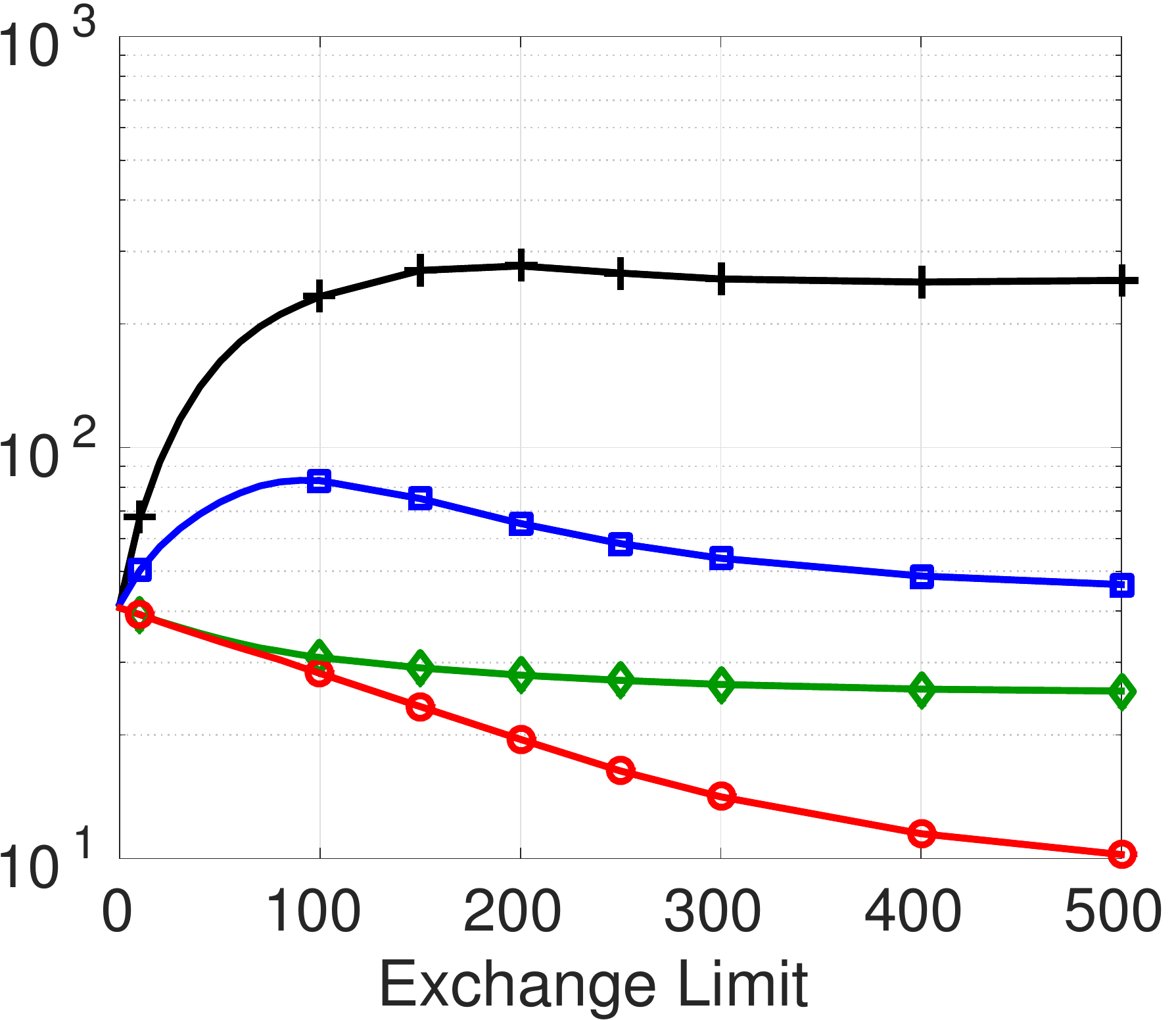}\label{fig:exchange_fourth}}
\caption{Effect of Exchange Limit on Performance Metrics}
\label{fig:res_exchange}
\end{figure}

Finally, we assume that the contact duration is limited so that the
number of messages during any meeting is restricted by an
{\em exchange limit}. We set the buffer size to
$1000$ messages so that all the algorithms reach their best possible
delivery rate. This 
buffer size implies $50$ MB of node memory if each
message is $50$ KB. We check the effect of varying the exchange limit on the
network performance. Figure \ref{fig:res_exchange} shows the four
comparison metrics for exchange limits in the interval $0$ to $500$
messages. As we see in the figure, MinLat cannot achieve the optimum
average latency for some values of exchange limit, but it still has
the best performance in terms of buffer occupancy.

\begin{figure}[!htb]
\centering
\subfloat[Absolute Estimated Errors]{\includegraphics[width=1.7 in]{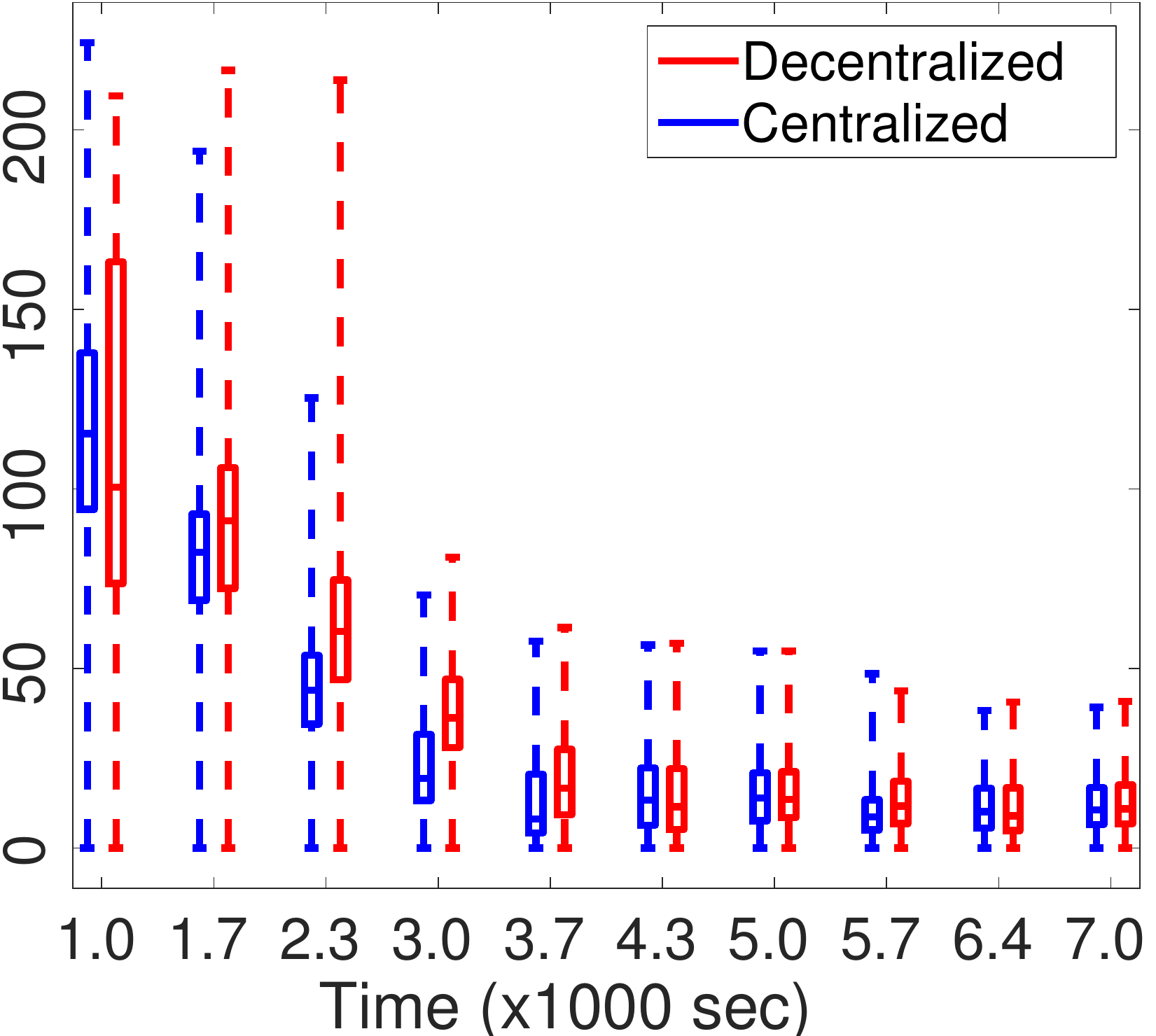}\label{fig:resultsA}}\hspace*{0.5em}
\subfloat[Absolute Achieved Errors]{\includegraphics[width=1.7 in]{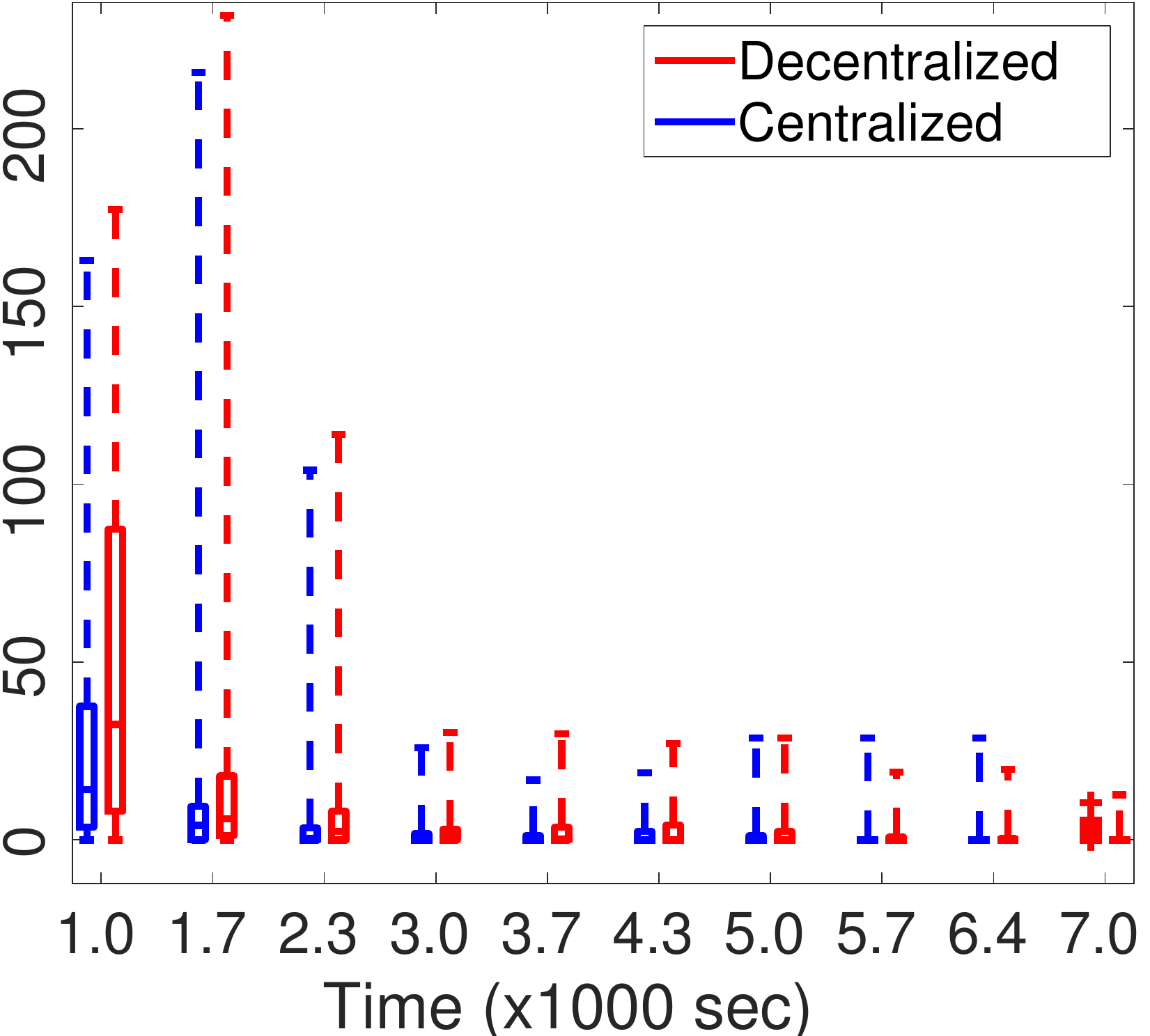}\label{fig:resultsB}}
\caption{Absolute Errors of Estimated and Achieved Latencies. The central mark of each box is the median, the edges of the box
are the 25th and 75th percentiles, the whiskers extend to the highest values not considered as outliers, and outliers are plotted
individually.}
\label{fig:res3}
\end{figure}

The simulation results displayed in Figures
\ref{fig:res_TTL}-\ref{fig:res_exchange} illustrate that although
MinLat has been designed for an ideal network model, it has an
acceptable performance when we impose realistic conditions such as finite TTL,
buffer size, and exchange limit.  
Our final set of simulations explores the impact of including the
recursive maximum likelihood estimation of meeting rates. We conduct
experiments using both the centralized and
decentralized algorithms operating on the extension of
network {\em I} to $100$ nodes. We select the $\lambda$ parameters of
intermeeting times from a uniform distribution $U[0,0.01]$. The
destination node is randomly chosen from the $N$ nodes of the network
based on a uniform distribution and is fixed throughout the
simulation. We run the simulation for $5\times 10^4$ seconds.
We examine the error terms in Theorem 4,
averaging over all nodes.  Figure \ref{fig:resultsA} shows the average
absolute difference between the estimated expected latencies and the
true minimum latencies, i.e.,
$\frac{1}{N}\sum_{k}|\widetilde{L}_{kd,t}(\widetilde{\mathbf{B}}_t,k) -
L_{kd}(\mathbf{B}^*)|$.  Figure~\ref{fig:resultsB} shows the
difference between the achieved average latencies and the true minimum latencies, i.e.,
$\frac{1}{N}\sum_{k}|L_{kd}(\widetilde{\mathbf{B}}_t) -
L_{kd}(\mathbf{B}^*)|.$
Figure \ref{fig:res3} indicates that the decentralized algorithm
achieves almost the same estimation error as the centralized
algorithm, suggesting that the limiting effect is the convergence
of the meeting rate estimates rather than the dissemination of latency
estimates through the network. As expected from Theorem \ref{theorem4}, we see that
the estimated and achieved errors both decay to zero as the time goes
by (for achieved latencies, it is almost zero for all nodes of the
network after $t=7000$ seconds).

Comparing Figures \ref{fig:resultsA} and \ref{fig:resultsB} shows that
for both centralized and decentralized scenarios, the average
difference between the achieved latency and the minimum latency is
less than the average difference between the estimated latency and the
minimum latency. This is expected, because the estimated latencies are
based on incorrect decision matrices and estimated meeting rates,
whereas the achieved latencies are derived from the actual meeting
rates. The results suggest that even when there remains substantial
inaccuracy in the expected latency estimates (e.g. at time $t=3000$
seconds), the algorithm identifies a close-to-optimal forwarding
decision matrix.

\section{Conclusion}\label{sec:conc}
In this paper, we used an analytical framework to model the
opportunistic data transfer among mobile devices in Delay Tolerant
Networks. In our model, the random intermeeting times of nodes are
assumed to be independent and exponentially distributed. We formulated
the routing/forwarding problem as an optimization problem in which the
goal is to minimize the sum of expected latencies from all nodes of
the network to a particular destination. We proved that the solution
of this problem is binary, i.e., when an arbitrary node meets any
other node in the network, its optimum forwarding rule dictates either
to always forward its messages to the encountered node or to never
forward any messages to it. We also showed that the solution of this
optimization problem minimizes the expected latency from each node of
the network to the destination as well. Based on these results, we
proposed centralized and distributed versions of an algorithm to find
the optimum forwarding decision rules and proved that each of these
algorithms result in the same solution. In order to evaluate the
performance efficiency of the suggested algorithms in different synthetic
and real-world networks, we chose four performance metrics as
comparison metrics and compared our proposed decentralized algorithm
(MinLat) with the most similar existing approaches. In order to
evaluate the performance of MinLat in more realistic scenarios, we
conducted simulations in larger networks with practical
constraints like limited message life (TTL), buffer size and message
exchange.

One of the main contributions of this work is relaxing the condition of
having complete knowledge of meeting rates at each node for making the
forwarding decisions. We used a recursive maximum likelihood procedure
(MinLat-E) to learn the meeting rates online and proved its
convergence in probability to the same optimal solution. The validity
of this theoretical result was assessed through
simulations. Moreover, we compared the convergence speed of the
proposed centralized and decentralized algorithms when the meeting
rates are estimated online. The simulation results show that the
decentralized algorithm has almost the same convergence rate as the
centralized algorithm, even though the network topology 
is not known at individual nodes. 

In future work, we aim to explore the effect of time-varying meeting
rates between nodes, which would motivate the use of filters to track
the rates. We also hope to examine whether it is possible to derive
similar expressions for expected latencies and optimal forwarding
rules for cases when the intermeeting times are not exponentially
distributed or are correlated.

 \appendices
\section{Proof of Lemma \ref{Lid}}\label{pl1}
\begin{proof}[\unskip\nopunct]When node $i$ commences in its routing of a packet, it must
  first wait a time $T_{w}$ before it meets one of its neighbors. The amount of time before node $i$ meets a specific neighbor
  $j$ is an exponentially distributed random variable with
  parameter $\lambda_{ij}$. The time $T_{w}$ is equal to
  the minimum of the exponentially distributed random variables
  corresponding to all neighbours $k \in \mathcal{S}_i$
  and its expected value is
\begin{equation}\label{Twait}
E(T_{w})=\frac{1}{\sum\limits_{k \in \mathcal{S}_i} \lambda_{ik}}
\end{equation}
The probability that $j$ is the first node that $i$
meets is $\frac{\lambda_{ij}}{\sum_{k \in \mathcal{S}_i}
  \lambda_{ik}}$. Hence $L_{id}(\bP)$ is
\begin{equation} \label{Lid_expansion} L_{id}(\bP)=E(T_{w})+\sum_{j
    \in \mathcal{S}_i}\frac{\lambda_{ij}}{\sum\limits_{k \in \mathcal{S}_i}
    \lambda_{ik}} [ p_{ij}L_{jd}(\bP)+(1-p_{ij})L_{id}(\bP)]
\end{equation}
The last term in \eqref{Lid_expansion} follows from the
memoryless property of the distributions. Subsituting \eqref{Twait} into
\eqref{Lid_expansion} leads to
\eqref{lem1}.
\end{proof}
\section{Proof of Theorem \ref{theorem1}}\label{pt1}
\begin{proof}[\unskip\nopunct]
  Let us assume that the nodes, excluding $d$, are labelled in
  ascending order of their expected latency under $\bP^*$, i.e.,
  $L_{1d}(\bP^*) < L_{2d}(\bP^*) < \dots < L_{N-1d}(\bP^*)$. For a
  given matrix $\bP$, we denote by $\bP_{\bar i}$ all rows of $\bP$
  except $i$. If we fix $\bp_i$ and $L_{kd}$ for $k\in\cS_i, k\neq j$
  for some $j\in \cS_i$, then $L_{id}$ is monotonically increasing
  with respect to $L_{jd}$ (see \eqref{lem1}). This implies that if we
  commence with any $\bP$ and change only $\bp_j$ to decrease
  $L_{jd}$, then all other $L_{id}$ such that $j \in \cS_i$ either
  decrease or remain the same. The matrix $\bP^*$ must therefore
  satisfy $\bp^*_j = \arg \min L_{jd}(\bP^*_{\bar j}, \bp_j)$ for all
  $j$. Otherwise we could choose an alternative $\bp^{\prime}_j$ that
  reduces $L_{jd}$ and hence achieves $\cU(\bP^{\prime})<\cU(\bP^*)$.
  
We can examine the partial derivative of $L_{id}$ with respect to
$p_{ij}$ at $\bP^\prime = (\bP_{\bar i}^*,\bp_i)$:
\begin{equation}
\frac{\partial L_{id}}{\partial p_{ij}}=\frac{\lambda_{ij}[\sum_{k\in \mathcal{S}_i}\lambda_{ik} p_{ik} (L_{jd}(\bP^\prime)-L_{kd}(\bP^\prime))-1]}{(\sum_{k\in \mathcal{S}_i}\lambda_{ik} p_{ik})^2}
\end{equation}
This derivative has the same sign as: $L_{jd}(\bP^\prime) -
\frac{1+\sum_{k\in \mathcal{S}_i}\lambda_{ik} p_{ik}
  L_{kd}(\bP^\prime)}{\sum_{k\in \mathcal{S}_i}\lambda_{ik} p_{ik}}$,
or equivalently $L_{jd}(\bP^\prime)-L_{id}(\bP^\prime)$.  This
expression for the derivative, together with the requirement that
$\bp^*_i = \arg \min L_{id}(\bP^*_{\bar i}, \bp_i)$, implies that
$\bp^*_{ij} = 0$ if $L_{id}(\bP^*)<L_{jd}(\bP^*)$ and $\bp^*_{ij} = 1$
if $L_{id}(\bP^*)>L_{jd}(\bP^*)$. Our assumption that the solution is
unique implies that $L_{id}(\bP^*)\neq L_{jd}(\bP^*)$. Otherwise, from
\eqref{Lid_expansion}, it is clear that we could choose any $p^*_{ij}$
between 0 and 1 and achieve the same $L_{id}(\bP^*)$, without
affecting any other $L_{jd}(\bP^*)$. This establishes
statement (1) of the theorem.

Although we have established that $L_{id}(\bP^*_{\bar i},\bp^*_i) = \min
L_{id}(\bP^*_{\bar i}, \bp_i)$, we have not yet shown that $\bP^*$
globally minimizes $L_{id}$. We establish this by
contradiction. Suppose $\mathbf{P}^*$ does not minimize the expected
latency for some non-empty set of nodes $\mathcal{N'} \subset
\mathcal{N}$. In other words, denoting the minimum expected latency
achieved via the minimization in \eqref{opt_node} for node $i$ by
$L_{id}^*$, we have
\begin{equation}
\forall i \in \mathcal{N}': \quad L_{id}^* < L_{id}(\mathbf{P}^*)
\end{equation}
Let node $s$ be the node in $\cN'$ such
that $L_{sd}^*<L_{kd}^*$ for all $k \in \cN', k \neq s$. Denote by
$\ell$ the ranking of the node with greatest expected latency under
$\bP^*$ such that $L_{\ell d}(\bP^*) < L_{sd}^*$. Based on the
discussion above, for each node $i \in \{1, 2,\dots,\ell\}$,
$p^*_{ik}=0$ for all $k>\ell$ and hence
$p^*_{is}=0$. Node $s$ must have at least one neighbour in the set $\{d,1,
2,\dots\ell\}$. Otherwise, it could not achieve an expected latency
under $\bP^*$ that is less than all nodes $\ell+1,\dots,N-1$ (observe from
 \eqref{lem1} that $L_{sd}(\bP^*) > \min_{p_{sj}>0} L_{jd}(\bP^*)$). 

 The matrix $\bP^\prime$ that achieves the minimum $L_{sd}^*$ must
 satisfy $p^\prime_{sk}=0$ for all $k\in \mathcal{N}'$, since for any
 matrix $\bP^\prime$ we have $L_{kd}(\bP') \geq L_{kd}^* >
 L_{sd}^*$. We also have $p^\prime_{sk} = 1$ for $k\in \cS_s\cap
 \{d,1,2,\dots,\ell\}$ if
 $L_{kd}(\bP^\prime)<L_{sd}(\bP^\prime)$. For a fixed choice of
 $\bp^\prime_s$ the value $L_{sd}(\bP^\prime)$ decreases if we can
 reduce $L_{kd}(\bP^\prime)$ for any $k$ such that $p^\prime_{sk} =
 1$. The matrix $\bP^*$ minimizes $L_{kd}$ for all $k\in
 \{1,2,\dots,\ell\}$, implying that $\bp^\prime_k = \bp^*_k$
 for all $k \in \{1,2,\dots,\ell\}$. Since $L_{kd}(\bP') =
 L_{kd}(\bP^*)$ for all $k \in \cS_s\cap \{d,1, 2,\dots,\ell\}$, it
 follows that $\bp^\prime_s = \bp^*_s$. For node $s$, the values
 of $\bp^{\prime}_j$ for $j \notin \{1,2,\dots,\ell, s\}$ have no
 impact on $L_{sd}$, so we have $L_{sd}(\bP^*) = L_{sd}(\bP^\prime) =
 L_{sd}^*$. This contradicts the original assumption that $\bP^*$ does
 not minimize the latency for all nodes $s \in \cN'$, and thus
 establishes statement (2) of the theorem.
\end{proof}
\section{Proof of Theorem \ref{theorem2}}\label{pt2}
\begin{proof}[\unskip\nopunct]
  We observe that for all $i \in \mathcal{N}$, $G_{id} \geq L_{id}^*$ (since the optimizations are
  the same). Based on Theorem 1 and its
  proof, the equality holds only if $j\in \mathcal{A}$ for all $j\in
  \mathcal{S}_{i}$ such that $L_{jd}^* < L_{id}^*$. The statements in
  the theorem follow based on an induction argument. 

  Suppose, without loss of generality, that the nodes are labelled in
  ascending order of expected latency under $\mathbf{B}^*$. For node $1$, the
  only neighbour with lower expected latency is the destination. In
  iteration 1, the destination is included in $\mathcal{A}$ and must
  be in $\mathcal{S}_1$. Recall that $L_{1d}^*>\min_{b^*_{1j}=1}
  L_{jd}^*$. Node 1 has the minimum expected latency according to the
  chosen labelling and Theorem 1, except for the destination
  itself. The relationship thus implies that $d\in \mathcal{S}_1$. We
  therefore have $G_{1d} = L_{1d}^* < L_{jd}^* \leq G_{jd}$, and node
  $1$ is selected to be added to $\mathcal{A}$, with $b_{1d} = 1$ and
  $b_{jd} = 0$ for all $j\neq d$. Statements 1a)-c) in the theorem
  clearly hold after one iteration, i.e., after the addition of node 1 to
  $\mathcal{A}$.

  Assume the same statements hold after the addition of node $k-1$ to
  $\mathcal{A}$. Then, for node $k$ we must have $j\in \mathcal{A}$
  for all $j\in \mathcal{S}_{k}$ such that $L_{jd}^* <
  L_{kd}^*$. Again this implies that $G_{kd}=L_{kd}^*<L_{jd}^* \leq
  G_{jd}$ for all $j>k$. Thus, node $k$ is correctly selected for
  addition to $\mathcal{A}$ and the statements 1a)-c) hold at the end
  of iteration $k$.

  It follows that the statements hold for all iterations of the
  algorithm, and after completion, when $\mathcal{A}=\mathcal{N}$, the
  second statement follows.
\end{proof}

\section{Proof of Theorem \ref{theorem3} and Proposition \ref{prop1}}\label{pt3}

\subsection{Proof of Theorem \ref{theorem3}}

\begin{proof}[\unskip\nopunct]
  Assume that the nodes are labelled in order of ascending expected
  latency under $\mathbf{B}^*$, i.e., $L_{1d}(\mathbf{B}^*) \leq
  L_{2d}(\mathbf{B}^*) \leq \dots \leq L_{N-1 d}(\mathbf{B}^*)$. Denote
  by $T_{1}$ the moment of time at which node $1$ meets the
  destination node. For $k=2,\dots,N$ denote by $T_{k}$ the earliest
  time by which node $k$ has met all nodes in the set
  $\{1,\dots,k-1\}\cap \mathcal{S}_k$ in the time period
  $(T_{k-1},T_{k}]$. Due to the assumption that the inter-meeting
  times are exponentially distributed, $T_{N}$ is finite with
  probability $1$.

  At $T_{1}$, node $1$ learns its meeting rate with the
  destination ($\lambda_{1d}$). Since the estimated latencies are initialized
  to $\infty$ and due to the update equations in Algorithm
  \ref{algo:DistLatMin}, the estimation that node $1$ has at
  $T_{1}$ of the latencies of its neighbors $i \in
  \mathcal{S}_{1}$ are upper-bounds, i.e. $\widehat{L}_{id}(1) \geq
  L_{id}(\mathbf{B}^*)$. As discussed in the proof of the previous
  theorems, the minimizer $\mathbf{b}^*_{1}$ has $b_{1d}=1$ and
  $b_{1j}=0$ for all $j\neq d$. At time $T_{1}$,
  since the term involving $d$ in the update equation of Algorithm
  \ref{algo:DistLatMin} has its minimum value, the vector
  $\mathbf{m}_{1}^*=\mathbf{b}_{1}^*$ identifies the same minimum
  latency $\widehat{L}_{1d}(1) = L_{1d}(\mathbf{B}^*)$. Hence,
  immediately after time $T_{1}$ we are guaranteed that
  $\mathbf{b}_{1}=\mathbf{b}_{1}^*$.

  At $T_{k}$, node $k$ is aware of the minimum expected latencies
  $\widehat{L}_{sd}(k)=L_{sd}(\mathbf{B^*})$ for the nodes in the set
  $\mathcal{V}_k = \{d,1,...,k\}\cap \mathcal{S}_{k}$. All other expected latencies
  are upper bounds, i.e. $\widehat{L}_{jd}(k) \geq L_{jd}(\mathbf{B}^*)$ for
  $j\notin \mathcal{V}_k$. The solution
  $\mathbf{b}^*_{k}$ takes value $1$ only for nodes in
  $\mathcal{V}_k$. The minimizer $\mathbf{m}^*_{k}$ at time $T_k$ is thus equal to
  $\mathbf{b}^*_{k}$ and achieves $\widehat{L}_{kd}(k)=L_{kd}(\mathbf{B}^*)$. Therefore, imediately
  after $T_{k}$ we will have $\mathbf{b}_{k}=\mathbf{b}_{k}^*$. This argument applies until just after $T_{N}$, at which point we
  have $\mathbf{B} = \mathbf{B} ^*$. Since $T_{N}$ is finite with
  probability $1$, the statement of the theorem follows.
  
\subsection{Proof of Proposition~\ref{prop1}}

Algorithm \ref{algo:DistLatMin} has converged when all of the nodes have
met all of their relay candidates and have identified their optimum
forwarding rules. Thus, $E(T_N)$, where $T_{N}$ defined above in the proof of
Theorem~\ref{theorem3}, is the expected convergence
time. Considering the worst case where an arbitrary node $k>1$
is connected to all the nodes in the set $\{1,...,k-1\}$, we have
\begin{equation}\label{Tk}
E(T_k-T_{k-1})=E(\underset{i \in \{1,...,k-1\}}{\max} x_i)
\end{equation}
where $x_i$ denotes the intermeeting time of node $k$ with its
neighbour $i$ and follows an exponential distribution with parameter
$\lambda_{ki}$. Using a standard result for the expected value of
the maximum of non-identical independent exponential random variables, we have
\begin{equation}
E(T_k-T_{k-1})=\sum_{i=1}^{k-1} \frac{1}{\lambda_{ki}}-\sum_{i=1}^{k-1} \sum_{j=i+1}^{k-1} \frac{1}{\lambda_{ki}+\lambda_{kj}}
+ \sum_{i=1}^{k-1} \sum_{j=i+1}^{k-1}\sum_{l=j+1}^{k-1} \frac{1}{\lambda_{ki}+\lambda_{kj}+\lambda_{kl}}-\dots
\end{equation}
An upper bound on this value is
$E(T_k-T_{k-1})<\sum_{i=1}^{k-1} \frac{1}{\lambda_{ki}}<\frac{k-1}{\underset{i}{\min} \lambda_{ki}}$
Therefore, an upperbound on the convergence time of Algorithm \ref{algo:DistLatMin} is $
E(T_N)<\frac{1}{\lambda_{1d}}+\sum_{l=2}^{N} \frac{l-1}{\underset{\underset{\lambda_{li}>0}{i \in \{1,\dots,l-1\}}}{\min} \lambda_{li}}$.
\end{proof}

\section{Proof of Lemma \ref{lem2}}\label{pl2}
\begin{proof}[\unskip\nopunct]
  Let's return to considering the probability decision variable vector
  $\bp$ instead of binary decision variable vector. (We still expect
  the optimal solution to be of the binary form). Without loss of
  generality, we relabel the nodes $k \in \cS_i$ by labels
  $1,...,|\cS_i|$. The optimization problem that node $i$ tries to
  solve in Algorithm \ref{algo:DistLatMin} follows this general form
  which is known as linear fractional optimization problems:
\begin{equation}\label{optLin}
\begin{aligned}
& \widehat{L}_{id}(i)=\underset{\mathbf{p}^i}{\text{min}}
& & \frac{\mathbf{c}^T \mathbf{p}^i +\alpha}{\mathbf{d}^T \mathbf{p}^i +\beta} \\
& \text{subject to}
& & \mathbf{A}\mathbf{p}^i\leq \mathbf{b}
\end{aligned}
\end{equation}
where $\mathbf{p}^i_{|\cS_i| \times 1}=[p_{i1},...,p_{i|\cS_i|}]^T$, $\alpha=1$,
$\mathbf{c}_{|\cS_i| \times 1}=[\lambda_{i1}\widehat{L}_{1d}(i),...,\lambda_{i|\cS_i|} \widehat{L}_{|\cS_i|d}(i)]^T$,  $\beta=0$, 
 $\mathbf{d}_{|\cS_i| \times 1}=[\lambda_{i1},...,\lambda_{i|\cS_i|}]^T$, $\mathbf{b}_{2|\cS_i| \times 1}=[1,...,1,0,...,0]^T$,and the elements of the matrix $\mathbf{A}_{2|\cS_i| \times |\cS_i|}$ are
\[
 \mathbf{A}_{ij} =
  \begin{cases}
   1 & \text{if } i<|\cS_i| \quad \& \quad j=i \\
   -1 & \text{if } i>|\cS_i| \quad \& \quad j=i-|\cS_i| \\
   0       & \text{otherwise }
  \end{cases}
\]
After applying the following parameter changes (the Charnes-Cooper transformation), we have
\begin{equation}
\bx=\frac{1}{\mathbf{d}^T\bp^i}\bp^i \qquad y=\frac{1}{\mathbf{d}^T\bp^i}
\end{equation}
and the optimization problem \ref{optLin} converts to
\begin{equation}
\begin{aligned}
& \underset{\mathbf{x},y}{\text{min}}
& & \mathbf{c}^T\mathbf{x}+\alpha y \\
& \text{subject to}
& & \mathbf{d}^T\mathbf{x}=1\\
& & & \mathbf{A}\mathbf{x} \leq \mathbf{b}y\\
& & & y\geq 0
\end{aligned}
\end{equation}
which is a linear optimization problem and can be solved using Linear Programming (LP) solution methods.
\end{proof}

\section{Proof of Theorem \ref{theorem4}}\label{pt4}
\begin{proof}[\unskip\nopunct]
  We first prove the statement of the theorem for estimated latencies.
  Without loss of generality, we relabel the nodes such that
  $L_{1d}(\mathbf{B}^*)<L_{2d}(\mathbf{B}^*)<...<L_{Nd}(\mathbf{B}^*)$. At
  node $i$, MinLat-E forms estimates of the meeting rates with the
  contact graph neighbours, $\widehat{\lambda}_{ik}$ for $k \in
  \mathcal{S}_i$, and the expected latencies from each neighbour
  $\widetilde{L}_{kd}(\widetilde{\mathbf{B}}_t,i)$. The result is a sequence of random
  variables $\{\widetilde{L}_{id}(\widetilde{\mathbf{B}}_t,i)\}$, with a variable being added
  to the sequence each time node $i$ meets another
  node. \cite{berk1972consistency} shows that the maximum likelihood
  estimator of the meeting rates is consistent, i.e., $\forall
  i,j\in\mathcal{N},\widehat{\lambda}_{ij} \xrightarrow{p}
  \lambda_{ij}$. More precisely,
\begin{equation}\label{consistency}
\forall \epsilon_{ij}>0 : \underset{t -> \infty}{\lim} P(|\hat{\lambda}_{ij}-\lambda_{ij}|< \epsilon_{ij})=1
\end{equation}
Equivalently, writing $\epsilon_{ij}=\epsilon_0 \lambda_{ij}$, we have
for any $\delta > 0$ and $\epsilon_{0}>0$, there exists a $t_0 >0$ such that for all $t>t_0$:
\begin{equation}
P((1-\epsilon_0)\lambda_{ij}<\widehat{\lambda}_{ij,t}<
(1+\epsilon_0)\lambda_{ij}) > (1-\delta) \label{prob_bound}
\end{equation}
where $\widehat{\lambda}_{ij,t}$ denotes the estimate of the meeting
rate between nodes $i$ and $j$ at time $t$.

We show that for any set of meeting rates
$\{\lambda_{ij}\}_{i,j\in\mathcal{N}}$, there exists an
$\epsilon=\epsilon_0$ for which the optimum forwarding decision matrix
in MinLat-E ($\widetilde{\mathbf{B}}_t$) is the same as the optimum
forwarding decision matrix in MinLat ($\mathbf{B}^*$) with desirably
high probability. In order to do so, we find upper and lower bounds
(that apply with high probability) on the estimated expected latencies
for the optimal decision matrices identified by both MinLat and
MinLat-E.  We first demonstrate a relationship between the estimated and true
expected latencies that would hold if the nodes employed the optimal
decision matrix $\mathbf{B}^*$. 
We show that there exists a $t_{\delta,i}$ such that for each node $i \in
\mathcal{N}$, with probability greater than $1-\delta$, for any
positive $\delta$, we have for
all $t>t_{\delta,i}$:
\begin{equation}\label{tighbound}
\frac{(1-\epsilon_0)^{i-1}}{(1+\epsilon_0)^i} L_{id}(\mathbf{B}^*) < \widetilde{L}_{id,t}(\mathbf{B}^*,i) < \frac{(1+\epsilon_0)^{i-1}}{(1-\epsilon_0)^i} L_{id}(\mathbf{B}^*)
\end{equation}
We derive \eqref{tighbound} by induction.  From the arguments made in
Theorem 3, we know that under $\mathbf{B}^*$ an arbitrary node $i$
will not forward any messages to a node that does not belong to the
set $\{1,..,i-1,d\}$. For $i=1$, after time $t_{\delta,1}$ such
that~\eqref{prob_bound} holds, we have, with probability greater than
$1-\delta$ for $t>t_{\delta,1}$:
\begin{equation}
\frac{L_{1d}(\mathbf{B}^*)}{1+\epsilon_0}=\frac{1}{1+\epsilon_0}\frac{1}{\lambda_{1d}}<\widetilde{L}_{1d,t}(\mathbf{B}^*,1)=\frac{1}{\widehat{\lambda}_{1d,t}}
<\frac{1}{1-\epsilon_0}\frac{1}{\lambda_{1d}}=\frac{L_{1d}(\mathbf{B}^*)}{1-\epsilon_0}
\end{equation}
Suppose \eqref{tighbound} holds for all the nodes
$1,\dots,k-1$. Denote by $t_{jk}>t$ the last meeting between node $j$
and $k$ that occurs subsequent to $t_{\delta,k-1}$ but prior to a
considered time $t$. Then we can identify a $t^u_{\delta,k}$ so that the following relationship holds
with probability greater than $1-\delta$ for $t>t^u_{\delta,k}$:
\begin{subequations}
\begin{align}
&\widetilde{L}_{kd,t}(\mathbf{B}^*,k)
=\frac{1}{\Sigma_{j\in \mathcal{S}_k}b_{kj}^*\widehat{\lambda}_{kj}}(1+\Sigma_{j\in \mathcal{S}_k}b_{kj}^*\widehat{\lambda}_{kj}\widetilde{L}_{jd,t_{jk}}(\mathbf{B}^*,j))\\
&<(\frac{1}{1-\epsilon_0}\frac{1}{\Sigma_{j\in \mathcal{S}_k}b_{kj}^*\lambda_{kj}})
(1+\Sigma_{j\in \mathcal{S}_k}b_{kj}^*(1+\epsilon_0)\frac{(1+\epsilon_0)^{k-2}}{(1-\epsilon_0)^{k-1}}L_{jd}(\mathbf{B}^*))\\
&<\frac{(1+\epsilon_0)^{k-1}}{(1-\epsilon_0)^{k}} L_{kd}(\mathbf{B}^*)
\end{align}
\end{subequations}
Here we have chosen $t^u_{\delta,k}$ to be sufficiently large such
that the inequality on the second line holds with probability
exceeding $1-\delta$.  Similarly we can identify a $t^\ell_{\delta,k}$
so that the lower bound in~\eqref{tighbound} holds with probability
greater than $1-\delta$ for $t>t^\ell_{\delta,k}$.  By taking
$t_{\delta,k} = \max\{t^\ell_{\delta,k},t^u_{\delta,k} \}$, we see
that \eqref{tighbound} holds for node $k$ as well, and by induction,
holds for all nodes $i \in \mathcal{N}$.

We now turn our attention to the forwarding matrix determined by
MinLat-E ($\widetilde{\mathbf{B}}_t$).  We first consider a scenario where the estimates of
the meeting rates are frozen after a certain time $t_f$. The
minimization procedure in MinLat-E is the same as in MinLat, but
operates on the estimates of the meeting rates. If these estimates are
held constant, then the results in Theorems 1-3 apply, with the
substitution of $\widehat{\lambda}_{ij}$ anywhere we make use of
$\lambda_{ij}$. With probability 1, the optimization algorithm will
thus converge after a finite time $t'$, and the estimated expected
latencies will be consistent across the network. Hence, there exists a
labeling $\{\hat{1},\hat{2},...\hat{N}\}$ for which
$\widetilde{L}_{\hat{1}d}(\widetilde{\mathbf{B}}_{t_f+t'},\hat{1}) <
\widetilde{L}_{\hat{2}d}(\widetilde{\mathbf{B}}_{t_f+t'},\hat{2}) <
\dots <
\widetilde{L}_{\hat{N}d}(\widetilde{\mathbf{B}}_{t_f+t'},\hat{N})$ We
can now employ the same argument that was used above for
$\mathbf{B}^*$ to determine that there is a finite time $t_{\delta,\hat{k}}$ such
that the following bound holds for all $\hat{k} \in \mathcal{N}$ with probability greater than
$1-\delta$ for all $t>t_{\delta,\hat{k}}$. 
\begin{equation}\label{lhat}
\begin{aligned}
\frac{(1-\epsilon_0)^{\hat{k}-1}}{(1+\epsilon_0)^{\hat{k}}} L_{\hat{k}d}(\widetilde{\mathbf{B}}_{t})<\widetilde{L}_{\hat{k}d}(\widetilde{\mathbf{B}}_{t},\hat{k})<\frac{(1+\epsilon_0)^{\hat{k}-1}}{(1-\epsilon_0)^{\hat{k}}} L_{\hat{k}d}(\widetilde{\mathbf{B}}_{t})
\end{aligned}
\end{equation}
In the actual MinLat-E algorithm, the meeting rates
$\widehat{\lambda}$ are not frozen after $t_0$, but continue to be
updated as more meetings occur. This only results in the probabilistic
bounds on $\widehat{\lambda}$ being tighter, and hence can only
tighten the bounds on ${L}_{\hat{k}d}(\widetilde{\mathbf{B}}_{t},\hat{k})$.

To avoid having node-specific bounds on the accuracy of the estimates,
we can rewrite the bounds as:
\begin{equation}
\begin{aligned}
\frac{(1-\epsilon_0)^{N-1}}{(1+\epsilon_0)^{N}} L_{kd}(\mathbf{B})<\widetilde{L}_{kd}(\mathbf{B},k)<\frac{(1+\epsilon_0)^{N-1}}{(1-\epsilon_0)^{N}} L_{kd}(\mathbf{B})
\end{aligned}
\end{equation}
This bound holds for both $\mathbf{B} = \mathbf{B}^*$ and
$\mathbf{B} = \widetilde{\mathbf{B}}_{t}$ with probability at least
$1-\delta_0$ after some time $t_0$. 

Our goal is to show that there exists a moment of time after which
$\widetilde{\mathbf{B}}_t=\mathbf{B}^*$ is true with desirably high
probability. We can accomplish this by showing that there exists an
$\epsilon_0$ (and thus an associated time $t_0$) for which the upper-bound on
$\widetilde{L}_{id}(\mathbf{B}^*,i)$ is less than the lower-bound on
$\widetilde{L}_{id}(\widetilde{\mathbf{B}}_t,i)$ for any $\widetilde{\mathbf{B}}_t \neq
\mathbf{B}^*$ for all $t>t_0$. If this is the case, then with
probability exceeding $1-\delta_0$, the optimization procedure that
derives $\widetilde{\mathbf{B}}_t$ will set it to $\mathbf{B}^*$,
because it minimizes the estimated latencies.  Hence, $\epsilon_0$ should satisfy
\begin{equation}
L_{id}(\mathbf{B}^*)<(\frac{1-\epsilon_0}{1+\epsilon_0})^{2N-1} L_{id}(\widetilde{\mathbf{B}}_t)\,\,,
\end{equation}
which leads to
\begin{equation}
\epsilon_0<\frac{e^{\frac{\ln(K)}{2N-1}}-1}{e^{\frac{\ln(K)}{2N-1}}+1}\,,
\end{equation}
where $K=\frac{\min_{\mathbf{B}\neq\mathbf{B}^*} L_{id}(\mathbf{B})}{L_{id}(\mathbf{B}^*)}$ .

Now that we have established that after a finite amount of time
$\widetilde{\mathbf{B}}_t=\mathbf{B}^*$ occurs with a probability desirably
close to one, we can show that
$\{\widetilde{L}_{id}(\widetilde{\mathbf{B}}_t,i)\} \xrightarrow{p} L_{id}(\mathbf{B}^*)$ for any $i\in \mathcal{N}$ using the following properties:
\begin{enumerate}
\item If $X_n \xrightarrow{p (\text{or } d)} X$, then $g(X_n) \xrightarrow{p (\text{or } d)} g(X)$ (Continuous Mapping Theorem); 
\item If $X_n \xrightarrow{p} X$, then $X_n \xrightarrow{d} X$;
\item If $X_n \xrightarrow{d} c\in \mathbb{R}$, then $X_n \xrightarrow{p} X$; 
\item If $X_n \xrightarrow{d} X$ and $Y_n \xrightarrow{d} c \in \mathbb{R}$, then $g(X_n,Y_n) \xrightarrow{d} g(X,c)$ (Slutsky's Theorem),
\end{enumerate}
where $\xrightarrow{d}$ denotes the convergence in distribution and
$g:\mathbb{R}\times\mathbb{R} \rightarrow \mathbb{R}$ is an arbitrary
continuous function.

Again consider the node labelling such that $L_{1d}(\mathbf{B}^*) <
L_{2d}(\mathbf{B}^*) < \dots < L_{Nd}(\mathbf{B}^*)$. For node $1$,
$\{\widetilde{L}_{1d}(\widetilde{\mathbf{B}}_t,1)=\frac{1}{\widehat{\lambda}_{1d}}\}\xrightarrow{p}
L_{1d}(\mathbf{B}^*)=\frac{1}{\lambda_{1d}}$ which is obvious from
property 1. For any node $k>1$, if
$\{\widetilde{L}_{jd}(\widetilde{\mathbf{B}}_t,j)\} \xrightarrow{p}
L_{jd}(\mathbf{B}^*)$ holds for any $j \in \{1,...,k-1\} $, then due
to properties 2-4 we have,
\begin{equation}\label{2conv}
\begin{aligned}
&\forall j \in \{1,...,k-1,d\}\cap \mathcal{S}_k: \\
&\quad \quad \{\widehat{\lambda}_{kj} \widetilde{L}_{jd}(\widetilde{\mathbf{B}}_t,k)\}\xrightarrow{d} \lambda_{kj} L_{jd}(\mathbf{B}^*) \in \mathbb{R}\,\,,\\
\end{aligned}
\end{equation}
and 
\begin{equation}\label{2conv2}
\{\Sigma_{j \in \{1,...,k-1,d\}} \hat{b}^*_{kj} \widehat{\lambda}_{kj}\} \xrightarrow{p} \Sigma_{j \in \{1,...,k-1,d\}} b^*_{kj} \lambda_{kj}\,\,.
\end{equation}
Property 1 in combination with \eqref{2conv} and \eqref{2conv2} results in the statement of the theorem.

For the achieved expected latencies, $L_{kd}(\widetilde{\mathbf{B}}_t)$, as opposed to those estimated at
the nodes, the proof is more straightforward. For a given decision
matrix, $\widetilde{\mathbf{B}}_t$, the expected latencies
$L_{kd}(\widetilde{\mathbf{B}}_t)$s are functions of the true
meeting rates $\lambda$ and are thus not random variables. Thus, the sequence
$\{{L}_{id}(\widetilde{\mathbf{B}}_t)\}$ is only a function of the random sequences $\{
\widehat{\lambda}_{ij} \}$, $j \in \mathcal{S}_i$ via the optimization
that determines $\widetilde{\mathbf{B}}_t$. Since we have
 established that $\widetilde{\mathbf{B}}_t$ converges in probability to
$\mathbf{B}^*$, it follows that $\{{L}_{id}(\widetilde{\mathbf{B}}_t)\}$ converges in
probability to $L_{id}(\mathbf{B}^*)$ due to property 1.

\end{proof}

\ifCLASSOPTIONcaptionsoff
  \newpage
\fi

\bibliographystyle{IEEEtran}
\bibliography{references}

\begin{thebibliography}{10}
\providecommand{\url}[1]{#1}
\csname url@samestyle\endcsname
\providecommand{\newblock}{\relax}
\providecommand{\bibinfo}[2]{#2}
\providecommand{\BIBentrySTDinterwordspacing}{\spaceskip=0pt\relax}
\providecommand{\BIBentryALTinterwordstretchfactor}{4}
\providecommand{\BIBentryALTinterwordspacing}{\spaceskip=\fontdimen2\font plus
\BIBentryALTinterwordstretchfactor\fontdimen3\font minus
  \fontdimen4\font\relax}
\providecommand{\BIBforeignlanguage}[2]{{%
\expandafter\ifx\csname l@#1\endcsname\relax
\typeout{** WARNING: IEEEtran.bst: No hyphenation pattern has been}%
\typeout{** loaded for the language `#1'. Using the pattern for}%
\typeout{** the default language instead.}%
\else
\language=\csname l@#1\endcsname
\fi
#2}}
\providecommand{\BIBdecl}{\relax}
\BIBdecl

\bibitem{voyiatzis2012survey}
A.~Voyiatzis, ``A survey of delay- and disruption-tolerant networking
  applications,'' \emph{J. Internet Engineering}, vol.~5, no.~1, 2012.

\bibitem{khabbaz2012disruption}
M.~J. Khabbaz, C.~M. Assi, and W.~F. Fawaz, ``Disruption-tolerant networking: A
  comprehensive survey on recent developments and persisting challenges,''
  \emph{in IEEE Commun. Surveys \& Tutorials}, vol.~14, no.~2, pp. 607--640,
  2012.

\bibitem{cao2013routing}
Y.~Cao and Z.~Sun, ``Routing in delay/disruption tolerant networks: A taxonomy,
  survey and challenges,'' \emph{in IEEE Commun. Surveys \& Tutorials},
  vol.~15, no.~2, pp. 654--677, 2013.

\bibitem{wei2014survey}
K.~Wei, X.~Liang, and K.~Xu, ``A survey of social-aware routing protocols in
  delay tolerant networks: applications, taxonomy and design-related issues,''
  \emph{in IEEE Commun. Surveys \& Tutorials}, vol.~16, no.~1, pp. 556--578,
  2014.

\bibitem{juang2002energy}
P.~Juang, H.~Oki, Y.~Wang, M.~Martonosi, L.~S. Peh, and D.~Rubenstein,
  ``Energy-efficient computing for wildlife tracking: Design tradeoffs and
  early experiences with {Z}ebra{N}et,'' in \emph{ACM Sigplan Notices},
  vol.~37, no.~10, 2002, pp. 96--107.

\bibitem{small2003shared}
T.~Small and Z.~J. Haas, ``The shared wireless infostation model: a new ad hoc
  networking paradigm (or where there is a whale, there is a way),'' in
  \emph{Proc. ACM Int. Symp. Mobile Ad hoc Net.\& Comput. (MobiHoc)},
  Annapolis, MD, Jun 2003, pp. 233--244.

\bibitem{pataki2014sensor}
B.~E. Pataki and L.~Kov{\'a}cs, ``Sensor data collection experiments with
  chaoster in the fed4fire federated testbeds,'' in \emph{IEEE Wireless and
  Mobile Comput., Net. and Comm. (WiMob)}, Larnaca, Cyprus, Oct. 2014.

\bibitem{pentland2004daknet}
A.~Pentland, R.~Fletcher, and A.~Hasson, ``Daknet: Rethinking connectivity in
  developing nations,'' \emph{IEEE Computer}, vol.~37, no.~1, pp. 78--83, Jan.
  2004.

\bibitem{han2012mobile}
B.~Han, P.~Hui, V.~A. Kumar, M.~V. Marathe, J.~Shao, and A.~Srinivasan,
  ``Mobile data offloading through opportunistic communications and social
  participation,'' \emph{IEEE Trans. Mobile Comput.}, vol.~11, no.~5, pp.
  821--834, May 2012.

\bibitem{vahdat2000epidemic}
A.~Vahdat and D.~Becker, ``Epidemic routing for partially connected ad hoc
  networks,'' Duke Univ., Durham, NC, USA, Tech. Rep., 2000.

\bibitem{grossglauser2001mobility}
M.~Grossglauser and D.~Tse, ``Mobility increases the capacity of ad-hoc
  wireless networks,'' in \emph{Proc. IEEE Infocom}, vol.~3, Anchorage, AL,
  USA, Apr. 2001, pp. 1360--1369.

\bibitem{spyropoulos2005spray}
T.~Spyropoulos, K.~Psounis, and C.~S. Raghavendra, ``Spray and wait: an
  efficient routing scheme for intermittently connected mobile networks,'' in
  \emph{Proc. ACM SIGCOMM Workshop on Delay-tolerant Networks}, Philadelphia,
  PA, USA, Aug. 2005, pp. 252--259.

\bibitem{davis2001wearable}
J.~A. Davis, A.~H. Fagg, and B.~N. Levine, ``Wearable computers as packet
  transport mechanisms in highly-partitioned ad-hoc networks,'' in \emph{Proc.
  IEEE Int. Symp. on Wearable Computers}, Zurich, Germany, Oct. 2001, pp.
  141--148.

\bibitem{lindgren2003probabilistic}
A.~Lindgren, A.~Doria, and O.~Schel{\'e}n, ``Probabilistic routing in
  intermittently connected networks,'' \emph{ACM SIGMOBILE Mobile Comput. and
  Commun. Rev.}, vol.~7, no.~3, pp. 19--20, Jul. 2003.

\bibitem{burgess2006maxprop}
J.~Burgess, B.~Gallagher, D.~Jensen, and B.~N. Levine, ``Maxprop: Routing for
  vehicle-based disruption-tolerant networks.'' in \emph{Proc. IEEE Infocom},
  vol.~6, Barcelona, Spain, Apr. 2006, pp. 1--11.

\bibitem{jones2007practical}
E.~P. Jones, L.~Li, J.~K. Schmidtke, and P.~A. Ward, ``Practical routing in
  delay-tolerant networks,'' \emph{IEEE Trans. Mobile Comp.}, vol.~6, no.~8,
  pp. 943--959, 2007.

\bibitem{daly2007social}
E.~M. Daly and M.~Haahr, ``Social network analysis for routing in disconnected
  delay-tolerant manets,'' in \emph{Proc. ACM Int. Symp. Mobile Ad hoc Net.
  Comput. (MobiHoc)}, Montreal, Canada, Sep. 2007, pp. 32--40.

\bibitem{hui2011bubble}
P.~Hui, J.~Crowcroft, and E.~Yoneki, ``Bubble rap: Social-based forwarding in
  delay-tolerant networks,'' \emph{IEEE Trans. Mobile Comput.}, vol.~10,
  no.~11, pp. 1576--1589, 2011.

\bibitem{sharma2013contact}
D.~A. Sharma and M.~Coates, ``Contact graph based routing in opportunistic
  networks,'' in \emph{Proc. IEEE Global Conf. Sig. and Info. Proc.
  (GlobalSIP)}.\hskip 1em plus 0.5em minus 0.4em\relax Austin, TX, USA: IEEE,
  Dec. 2013.

\bibitem{xiao2013community}
M.~Xiao, J.~Wu, and L.~Huang, ``Community-aware opportunistic routing in mobile
  social networks,'' \emph{IEEE Trans. Computers}, vol.~63, no.~7, pp.
  1682--1695, 2013.

\bibitem{li2011impact}
Y.~Li, G.~Su, D.~O. Wu, D.~Jin, L.~Su, and L.~Zeng, ``The impact of node
  selfishness on multicasting in delay tolerant networks,'' \emph{IEEE Trans.
  Veh. Tech.}, vol.~60, no.~5, pp. 2224--2238, 2011.

\bibitem{sermpezis2014understanding}
P.~Sermpezis and T.~Spyropoulos, ``Understanding the effects of social
  selfishness on the performance of heterogeneous opportunistic networks,''
  \emph{Computer Commun.}, vol.~48, pp. 71--83, 2014.

\bibitem{zhang2013gossip}
H.~Zhang, Z.~Zhang, and H.~Dai, ``Gossip-based information spreading in mobile
  networks,'' \emph{IEEE Trans. Wireless Commun.}, vol.~12, no.~11, pp.
  5918--5928, 2013.

\bibitem{conan2008fixed}
V.~Conan, J.~Leguay, and T.~Friedman, ``Fixed point opportunistic routing in
  delay tolerant networks,'' \emph{IEEE J. Sel. Areas in Commun.}, vol.~26,
  no.~5, pp. 773--782, 2008.

\bibitem{xiao2013tour}
M.~Xiao, J.~Wu, C.~Liu, and L.~Huang, ``Tour: Time-sensitive opportunistic
  utility-based routing in delay tolerant networks,'' in \emph{Proc. IEEE
  Infocom}, Turin, Italy, Apr. 2013, pp. 2085--2091.

\bibitem{boldrini2010modelling}
C.~Boldrini, M.~Conti, and A.~Passarella, ``Modelling social-aware forwarding
  in opportunistic networks,'' in \emph{Proc. PERFORM (LNCS 6821)}.\hskip 1em
  plus 0.5em minus 0.4em\relax Vienna, Austria: Springer-Verlag, Oct. 2010, pp.
  141--152.

\bibitem{boldrini2012performance}
------, ``Performance modelling of opportunistic forwarding with imprecise
  knowledge,'' in \emph{Proc. Int. Symp. Model. and Opt. in Mobile, Ad Hoc and
  Wireless Net. (WiOpt)}, 2012, pp. 216--223.

\bibitem{ramanathan2007prioritized}
R.~Ramanathan, R.~Hansen, P.~Basu, R.~Rosales-Hain, and R.~Krishnan,
  ``Prioritized epidemic routing for opportunistic networks,'' in \emph{Proc.
  ACM Int. Workshop on Mobile Opportunistic Networks}, San Juan, Puerto Rico,
  Jun. 2007, pp. 62--66.

\bibitem{khouzani2012optimal}
M.~Khouzani, S.~Eshghi, S.~Sarkar, N.~B. Shroff, and S.~S. Venkatesh, ``Optimal
  energy-aware epidemic routing in {DTN}s,'' in \emph{Proc. ACM Int. Symp. on
  Mobile Ad hoc Net. \& Comput. (MobiHoc)}, South Carolina, USA, Jun. 2012, pp.
  175--182.

\bibitem{klein2010reaction}
D.~J. Klein, J.~Hespanha, and U.~Madhow, ``A reaction-diffusion model for
  epidemic routing in sparsely connected {MANET}s,'' in \emph{Proc. IEEE
  Infocom}, San Diego, USA, Mar. 2010, pp. 1--9.

\bibitem{wang2012analytical}
Q.~Wang and Z.~J. Haas, ``Analytical model of epidemic routing for
  delay-tolerant networks,'' in \emph{Proc. ACM Int. Workshop on High Perf.
  Mobile Opportunistic Sys.}, Paphos, Cyprus, Oct. 2012, pp. 1--8.

\bibitem{grasic2011evolution}
S.~Grasic, E.~Davies, A.~Lindgren, and A.~Doria, ``The evolution of a dtn
  routing protocol-prophetv2,'' in \emph{Proc. ACM Workshop on Challenged
  Networks}, Las Vegas, NV, USA, Sept. 2011, pp. 27--30.

\bibitem{shaghaghian2014opportunistic}
S.~Shaghaghian and M.~Coates, ``Opportunistic networks: Minimizing expected
  latency,'' in \emph{IEEE Wireless and Mobile Comput., Net. and Comm.
  (WiMob)}, Larnaca, Cyprus, Oct. 2014.

\bibitem{cai2009}
H.~Cai and D.~Y. Eun, ``Crossing over the bounded domain: from exponential to
  power-law intermeeting time in mobile ad hoc networks,'' \emph{IEEE/ACM
  Transactions on Networking}, vol.~17, no.~5, pp. 1578--1591, Oct. 2009.

\bibitem{chaintreau2007impact}
A.~Chaintreau, P.~Hui, J.~Crowcroft, C.~Diot, R.~Gass, and J.~Scott, ``Impact
  of human mobility on opportunistic forwarding algorithms,'' \emph{IEEE Trans.
  Mob. Comput.}, vol.~6, no.~6, pp. 606--620, Jun. 2007.

\bibitem{conan2007}
V.~Conan, J.~Leguay, and T.~Friedman, ``The heterogeneity of inter-contact time
  distributions: its importance for routing in delay tolerant networks,'' 2007,
  arXiv:cs/0609068v2 [cs.NI], LIP6.

\bibitem{gao2009}
W.~Gao, Q.~Li, B.~Zhao, and G.~Cao, ``Multicasting in delay tolerant networks:
  A social network perspective,'' in \emph{Proc. ACM MobiHoc}, New Orleans, LA,
  USA, May 2009, pp. 299--308.

\bibitem{lee2010}
K.~Lee, Y.~Y. amd J.~Jeong, H.~Won, I.~Rhee, and S.~Chong, ``Max-contribution:
  On optimal resource allocation in delay tolerant networks,'' in \emph{Proc.
  IEEE Infocom}, San Diego, CA, USA, Mar. 2010, pp. 1--9.

\bibitem{zhu2010}
H.~Zhu, L.~Fu, G.~Xue, Y.~Zhu, M.~Li, and L.~M. Ni, ``Recognizing exponential
  inter-contact time in {VANET}s,'' in \emph{Proc. IEEE Infocom}, San Diego,
  CA, USA, Mar. 2010, pp. 1--5.

\bibitem{cambridge-haggle-2006-01-31}
J.~Scott, R.~Gass, J.~Crowcroft, P.~Hui, C.~Diot, and A.~Chaintreau,
  ``{CRAWDAD} data set cambridge/haggle (v. 2006-01-31),'' Downloaded from
  http://crawdad.org/cambridge/haggle/, Jan. 2006.

\bibitem{berk1972consistency}
R.~H. Berk, ``Consistency and asymptotic normality of {MLE}'s for exponential
  models,'' \emph{The Annals of Math. Stat.}, vol.~43, no.~1, pp. 193--204,
  1972.

\end{thebibliography}

\end{document}